\documentstyle[11pt,aaspp4]{article}
\tighten
\eqsecnum

\begin{document}
%
% some useful Tex definitions..
%\def\et al. {{\it et al.} }
%\def\eg {{\it e.g.} }
%\def\ie {{\it i.e.} }
\def\kms {km~s$^{-1} \,$}
\def\kmsmpc {km~s$^{-1}$ Mpc$^{-1} \,$}
\def\lsim{ \lower .75ex \hbox{$\sim$} \llap{\raise .27ex \hbox{$<$}} }
\def\gsim{ \lower .75ex \hbox{$\sim$} \llap{\raise .27ex \hbox{$>$}} }
\def\items{\hangindent=0.5truecm \hangafter=1 \noindent}
\title{A Universal Density Profile from Hierarchical Clustering}

\author{Julio F. Navarro \altaffilmark{1}}
\affil{Steward Observatory, University of Arizona, Tucson, AZ, 85721, USA.}
\author{Carlos S. Frenk \altaffilmark{2}}
\affil{Physics Department, University of Durham, Durham DH1 3LE, England.}

\author{Simon D.M. White \altaffilmark{3}}
\affil{Max Planck Institut f\"ur Astrophysik, Karl-Schwarzschild
Strasse 1, D-85740, Garching, Germany.}

% Notice that each of these authors has alternate affiliations, which
% are identified by the \altaffilmark after each name.  The actual alternate
% affiliation information is typeset in footnotes at the bottom of the
% first page, and the text itself is specified in \altaffiltext commands.
% There is a separate \altaffiltext for each alternate affiliation
% indicated above.

\altaffiltext{1}{Bart J. Bok Fellow. E-mail: jnavarro@as.arizona.edu} 
\altaffiltext{2}{E-mail: C.S.Frenk@durham.ac.uk} 
\altaffiltext{3}{E-mail: swhite@mpa-garching.mpg.de} 

% The abstract environment prints out the receipt and acceptance dates
% if they are relevant for the journal style.  For the aasms style, they
% will print out as horizontal rules for the editorial staff to type
% on, so long as the author does not include \received and \accepted
% commands.  This should not be done, since \received and \accepted dates
% are not known to the author.

\begin{abstract}
We use high-resolution N-body simulations to study the equilibrium
density profiles of dark matter halos in hierarchically clustering
universes. We find that all such profiles have the same shape,
independent of halo mass, of initial density fluctuation spectrum, and
of the values of the cosmological parameters.  Spherically averaged
equilibrium profiles are well fit over two decades in radius by a
simple formula originally proposed to describe the structure of galaxy
clusters in a cold dark matter universe. In any particular cosmology
the two scale parameters of the fit, the halo mass and its
characteristic density, are strongly correlated. Low-mass halos are
significantly denser than more massive systems, a correlation which
reflects the higher collapse redshift of small halos. The
characteristic density of an equilibrium halo is proportional to the
density of the universe at the time it was assembled. A suitable
definition of this assembly time allows the same proportionality
constant to be used for all the cosmologies that we have tested. We
compare our results to previous work on halo density profiles and show
that there is good agreement. We also provide a step-by-step analytic
procedure, based on the Press-Schechter formalism, which allows
accurate equilibrium profiles to be calculated as a function of mass
in any hierarchical model.
\end{abstract}

%\keywords{globular clusters,peanut clusters,bosons,bozos}

% That's it for the front matter.  On to the main body of the paper.
% We'll only put in tutorial remarks at the beginning of each section
% so you can see entire sections together.
%
% In the first two sections, you should notice the use of the LaTeX \cite
% command to identify citations.  The citations are tied to the
% reference list via symbolic tags.  We have chosen the first three
% characters of the first author's name plus the last two numeral of the
% year of publication.  The corresponding reference has a \bibitem
% command in the reference list below.
%
% Please go to the LaTeX manual for a complete description of the
% \cite-\bibitem mechanism.
% citation example (\cite{hen61,lyn68,spi85})

\section{Introduction} \label{intro}
%The connection between cosmology and the structure of dark matter halos
%has received close attention ever since Gunn \& Gott (1972) showed
%that the structure of systems formed through gravitational collapse in
%an expanding universe may contain clues to the cosmological
%parameters. This link is particularly interesting because a growing
%number of observations can now probe in detail the mass profiles of
%dark halos surrounding galaxies and galaxy clusters. 

It is twenty-five years since the discovery that galaxies are
surrounded by extended massive halos of dark matter. A variety of
observational probes -- disk rotation curves, stellar kinematics, gas
rings, motions of globular clusters, planetary nebulae and satellite
galaxies, hot gaseous atmospheres, gravitational lensing effects --
are now making it possible to map halo mass distributions in some
detail. These distributions are intimately linked to the nature of the
dark matter, to the way halos formed, and to the cosmological context
of halo formation.

%Understanding the structure of dark matter halos and its dependence on
%global cosmological parameters is a fundamental problem which has attracted
%a good deal of attention since the original work of Gunn \& Gott
%(1972). Considerable progress has been achieved over the past twenty-five
%years, both theoretically and observationally. As a result it is now
%possible to test theoretical ideas directly against a a growing number of
%observations which probe in detail the mass profiles of the dark halos
%surrounding galaxies and galaxy clusters.  For example, high quality
%rotation curves of disk galaxies impose strict limits on the degree of
%central concentration of galactic dark matter (Persic \& Salucci 1991,
%Flores et al. 1993, Flores \& Primack 1994, Moore 1994), while studies of
%the dynamics of satellite galaxies yield estimates of the total mass and
%extent of dark halos (Zaritsky et al. 1993, Zaritsky \& White
%1994). Gravitational lensing effects in the strong and weak regimes can be
%combined with dynamical and X-ray studies to infer the shape and depth of
%the gravitational potential of galaxies and galaxy clusters (see, e.g.,
%Squires et al. 1996 and references therein). Even in elliptical galaxies,
%studies of absorption line profiles at large radii and of the X-ray
%emission from hot gas coronae place strong constraints on the mass profile
%of their surrounding halos (Carollo et al. 1995, Fabbiano 1989). These
%observations suggest that the dependence of halo structure on cosmology may
%be a useful tool to discriminate between competing cosmological models.

Insight into these links came first from analytic studies. Building on
the early work of Gunn \& Gott (1972), similarity solutions were
obtained by Fillmore \& Goldreich (1984) and Bertschinger (1985) for
the self-similar collapse of spherical perturbations in an Einstein-de
Sitter universe. Such solutions necessarily resemble power laws in the
virialized regions. Hoffman \& Shaham (1985) and Hoffman (1988)
extended this analysis by considering open universes, and by modeling
as scale-free spherical perturbations the objects which form by
hierarchical clustering from power-law initial density perturbation
spectra ($P(k) \propto k^n$).  They argued that isothermal structure
($\rho\propto r^{-2}$) should be expected in an Einstein-de Sitter
universe if $n\leq -2$, and that steeper profiles should be expected
for larger $n$ and in open universes.

%Insight into this dependence was first gained through analytic
%studies. Gunn \& Gott (1972) first recognized that the structure of halos
%formed through gravitational collapse in the expanding universe may be
%directly related to cosmological parameters.  Fillmore \& Goldreich (1984)
%and Bertschinger (1985) derived similarity solutions for the spherical
%collapse of scale-free perturbations in an Einstein-de Sitter
%($\Omega_0=1$) universe and concluded that the equilibrium mass profiles of
%dark matter halos are well approximated by power-laws. Hoffman \& Shaham
%(1985) and Hoffman (1988) extended this analysis to include open ($\Omega_0
%<1$) universes and power-law initial perturbation spectra ($P(k) \propto
%k^n$), and argued that the slope of the density profiles depends
%sensitively on the value of $n$ and $\Omega_0$. In particular, they
%obtained nearly isothermal profiles for $n \approx -2$ and $\Omega_0=1$ and
%gradually steeper profiles for larger values of $n$ and for lower values of
%$\Omega_0$.

Despite the schematic nature of these arguments, their general
predictions were verified as numerical data became available from
$N$-body simulations of hierarchical cosmologies. Power-law fits to
halo density profiles in a variety of simulations all showed a clear
steepening as $n$ increases or the density of the universe decreases
(Frenk et al. 1985, 1988; Quinn et al. 1986; Efstathiou et al. 1988;
Zurek, Quinn \& Salmon 1988; Warren et al. 1992; Crone, Evrard \&
Richstone 1994). An apparent exception was the work of West et
al. (1987), who found that {\it galaxy cluster} density profiles show
no clear dependence on $n$.

%The reason for this discrepant result has
%yet to be identified but one clue may come from the limited radial extent
%of the profiles' power-law behaviour (less than a decade in radius even in
%the best case; see, eg., Figure 3b of Crone et al. 1994). Slopes derived
%from power-law fits are sensitive to the radial range used in the fits, an
%effect that could, in principle, explain the disagreement between West et
%al (1987) and other work.

Significant departures from power-law behaviour were first reported by
Frenk et al. (1988), who noted that halo profiles in cold dark matter
(CDM) simulations steepen progressively with increasing
radius. Efstathiou et al. (1988) found similar departures -- at odds
with the analytic predictions -- in their simulations of scale-free
hierarchical clustering. They also noted that these departures were
most obvious in their best resolved halos. Similar effects were noted
by Dubinski \& Carlberg (1991) in a high resolution simulation of a
galaxy-sized CDM halo.  These authors found their halo to be well
described by a density profile with a gently changing logarithmic
slope, specifically that proposed by Hernquist (1990).

In earlier papers of this series we used high-resolution simulations
to study the formation of CDM halos with masses spanning about four
orders of magnitude, ranging from dwarf galaxy halos to those of rich
galaxy clusters (Navarro, Frenk \& White 1995, 1996).
%[Throughout this
%paper, we always measure halo masses, $M_{200}$, within the virial radius,
%$r_{200}$, defined as the radius of a sphere of mean interior density equal
%to $200 \, \rho_{crit}$, where $\rho_{crit}=3H^2/8 \pi G$ is the critical
%density required for closure and $H$ is Hubble's constant, whose value at
%$z=0$ we will denote by $H_0=100 \, h$ \kmsmpc. Halo circular speeds,
%$V_{200}=(GM_{200}/r_{200})^{1/2}$, are also measured at this radius unless
%otherwise specified. Numerical experiments show that in the Einstein-de
%Sitter universe this radius approximately delineates the boundary between 
%the virialized region of a system and the region where infall
%dominates (Cole \& Lacey 1996). For simplicity we shall continue to use
%this radius even when $\Omega_0 \ne 1$, although in this case spherical
%collapse models suggest that the virialized region may extend beyond
%$r_{200}$ (Eke, Cole \& Frenk 1996).]  
This work showed that the equilibrium density profiles of CDM halos of
all masses can be accurately fit over two decades in radius by the
simple formula,
$$ {\rho(r) \over \rho_{crit}}= {\delta_c \over (r/r_s)(1+r/r_s)^2}, 
\eqno(1) 
$$
where $r_s$ is a scale radius, $\delta_c$ is a characteristic
(dimensionless) density, and $\rho_{crit}=3H^2/8 \pi G$ is the
critical density for closure.  This profile differs from the Hernquist
(1990) model only in its asymptotic behaviour at $r \gg r_s$ (it tends
to $r^{-3}$ instead of $r^{-4}$). Power-law fits over a restricted
radial range have slopes which depend on the range fitted, steepening
from $-1$ near the center to $-3$ at large $r/r_s$.

This similarity between CDM halos of widely differing mass
is surprising in view of the strong dependence on power spectrum 
shape reported in earlier studies. The effective slope of the CDM 
power spectrum varies from $n_{\rm eff} \approx -2$ on galactic 
scales to $n_{\rm eff} \approx -1$ on cluster scales, so one might 
have expected shallower profiles in galaxy halos than in clusters. 
In fact, the opposite is true; low-mass halos are denser, i.e. have 
higher values of $\delta_c$, than high-mass halos. This property 
reflects the higher collapse redshift of the smaller systems. 
Power-law fits actually yield steeper slopes for less massive halos 
both when carried out at a fixed radius and when carried out at a 
fixed fraction of the virial radius
(see Figures 3 and 4 of Navarro, Frenk \& White 1996).

We argue below that the apparent relationship between profile shape
and initial power spectrum seen in earlier work results from
systematic differences in the characteristic density of the halos
chosen when comparing different models.  This interpretation is
reinforced by the recent work of Cole \& Lacey (1996) and Tormen,
Bouchet \& White (1996), who find that the profiles of massive halos
formed from power-law spectra are well described by eq.~1. They also
confirm the strong correlation between $\delta_c$ and halo mass seen
in our earlier work, and they find that, at a given mass, halos are
denser when $n$ is larger. Thus, the spectral index $n$ seems to
control the exact relationship between characteristic density and halo
mass, rather than the effective slope of the density profile.

It is clear from this discussion that a comprehensive study of this
problem needs to
consider the role of at least three factors:  the halo mass, the
power spectrum of initial density fluctuations, and the values of the
cosmological parameters. In this paper we present the results of a large
set of N-body simulations specifically designed to address these issues.
We consider a variety of hierarchical clustering models, including CDM and
power-law initial fluctuation spectra, as well as different values of the 
cosmological parameters
$\Omega_0$ and $\Lambda$. In each cosmology we study halos spanning a large
range in mass, carefully choosing numerical parameters so that all
systems are simulated with comparable numerical resolution.

The plan of the paper is as follows. Our numerical experiments are
described in \S 2. In \S 3 we present our results and in \S 4 we discuss
them in the context of earlier work. In \S 5 we summarize our main 
conclusions. An appendix lays out the formulae necessary to calculate
analytically the density profile of an equilibrium halo of any mass
in any hierarchical cosmology.

\section{The Numerical Experiments} \label{numexp}

\subsection{The Cosmological Models}

We analyze the structure of dark matter halos in 8 different
cosmologies. Five are Einstein-de Sitter ($\Omega_0=1$) models with
various power spectra: the standard biased CDM spectrum (SCDM model:
$\Omega_0=1$, $h=0.5$, $\sigma_8=0.63$) and four power-law or
``scale-free'' spectra with indices $n=0$, $-0.5$, $-1$, and
$-1.5$. Two further models also have power-law spectra ($n=0$ and
$-1$) but in an open universe ($\Omega_0=0.1$). The last model we
consider is a low-density CDM model with a flat geometry
(CDM$\Lambda$: $\Omega_0=0.25$, $\Lambda=0.75$, $h=0.75$,
$\sigma_8=1.3$). (Here and throughout this paper we express the
cosmological constant $\Lambda$ in units of $3 H_0^2$, so that a
low-density universe with a flat geometry has $\Omega_0+\Lambda=1$. We
also adopt the standard convention of writing the present Hubble
constant as $H_0 = 100~h~$km s$^{-1}$ Mpc$^{-1}$.)

The normalization of the CDM models is specified by $\sigma_8$, the
{\sl rms} mass fluctuation in spheres of radius $8 \, h^{-1}$
Mpc. Because of self-similarity, the normalization of the scale-free
models is arbitrary.  The evolutionary state of these models may be
fully specified by a single parameter, the current value of the
``nonlinear mass'', $M_{\star}(z)$. This mass scale is defined by
requiring that the variance of the linear overdensity field at
redshift $z=0$, smoothed with a top-hat filter enclosing a mass
$M=M_{\star}$, should equal the square of the critical density
threshold for spherical collapse by redshift $z$:
$\Delta_0^2(M_{\star}(z))= \delta_{crit}^2(z,\Omega_0,\Lambda)$. (See
the Appendix for details on the computation of $\delta_{crit}$). This
definition provides a ``natural'' way to scale the scale-free
simulations to physical units and to compare different cosmological
models. In scale-free models the mass scale defined by $M_{\star}(z)$
is the only physical scale, and therefore the structure of halos can
depend on mass only through the ratio $M/M_{\star}$.

%The critical density threshold, $\delta_{crit}(z)$, is the linear density
%contrast required for an isolated, overdense, spherical region to collapse.
%It is a particularly simple function of redshift in an Einstein-de Sitter
%universe, $\delta_{crit}(z)= 1.686 \, (1+z)$, but can be computed for any
%cosmology by considering the growth and collapse of an overdense spherical
%region. [See Lacey \&, Cole (1993) and Eke, Cole \& Frenk (1996) for
%derivations of the appropriate formulae, which are summarized in the
%Appendix.] At $z=0$, $\delta_{crit}$ depends only weakly on the
%cosmological parameters, varying from $1.69$ for an Einstein-de Sitter
%universe, to $1.67$ for a low-density model with a flat geometry
%($\Omega_0=0.1$, $\Lambda=0.9$), and to $1.62$ for an open model with
%$\Omega_0=0.1$ ($\Lambda=0$).

\subsection{The Simulations}

Large cosmological N--body simulations are required to simulate the
evolution of dark matter halos in their full cosmological context. However,
such simulations are not generally well suited to explore a large range of
halo masses. This is because systems of differing mass formed in a single
simulation are resolved to differing degrees. More massive systems are better
resolved because they contain more particles and because the gravitational
softening is a smaller fraction of the virial radius. These systematic
differences can introduce insidious numerical artifacts in the mass trends
that we wish to investigate. We circumvent this problem by using the
procedure outlined by Navarro, Frenk \& White (1996). Halos are first
identified in cosmological N-body simulations of large periodic boxes and
then resimulated individually at higher resolution. During the
resimulation the remainder of the original simulation is treated only 
to the accuracy needed to model tidal effects on the halo of interest.
The advantage of this procedure is that numerical parameters can be
tuned so that all halos are simulated with comparable resolution. Its
main disadvantage is that only one halo is modeled per simulation, so
that many simulations are needed to compile a representative halo sample.

\subsubsection{The cosmological simulations} 

The cosmological simulations were carried out using the P$^3$M code of
Efstathiou et al. (1985). The desired initial power spectrum was
generated by using the Zel'dovich approximation to displace particles
from a uniform initial load. The uniform load we used was either a
``glass'' configuration (White 1996) or a cubic grid. For simulations
with power-law power spectra, the amplitude of the initial
displacements was chosen by setting the power of the perturbed density
field to the white-noise level at the Nyquist frequency of the
particle grid. These simulations followed $10^6$ particles on a
$128^3$ mesh and were stopped when the nonlinear mass, $M_{\star}$,
corresponded to $1,000$-$2,000$ particles.  We identify this time with
the present ($z=0$). Since clustering evolves faster for more negative
values of $n$, an expansion factor of $9.5$ was sufficient for
$n=-1.5$ whereas an expansion factor of 90 was required for $n=0$
($\Omega_0=1$). Open models require even longer integrations, an
expansion factor exceeding 150 for $n=0$ and $\Omega_0=0.1$. These
simulations used the time-stepping and numerical scheme described in
Efstathiou et al (1988). (Note, however, that the definition of
$M_{\star}$ in that paper differs from the one we use here.)

We ran two P$^3$M simulations for each of our CDM models. The
SCDM runs followed $64^3$ particles in periodic boxes of $180 \, h^{-1}$
and $15 \, h^{-1}$ Mpc and were stopped when $\sigma_8=0.63$ ($M_{\star}
\approx 1.6 \times 10^{13} h^{-1} \, M_{\odot}$), the time which we
identify with the present. The CDM$\Lambda$ runs followed $10^6$ particles
in boxes of $140 \, h^{-1}$ and $46.67 \, h^{-1}$ Mpc, until $\sigma_8=1.3$
($M_{\star} \approx 4.1 \times 10^{13} h^{-1} \, M_{\odot}$).

\subsubsection{The individual halo simulations}

Halos to be resimulated at higher resolution were selected randomly at
$z=0$ from a list of clumps identified using a friends-of-friends
group finder with linking length set to $10 \%$ of the mean
interparticle separation. We chose masses in the range $0.1$--$10
M_{\star}$ for the power-law models and $0.01$--$100 M_{\star}$ for
the CDM models. (The mass range is larger for the CDM models because
in this case halos were chosen from two parent cosmological
simulations with different box sizes.)  Because we are interested the
structure of {\it equilibrium} halos we were careful not to choose for
analysis any halo which is far from virial equilibrium.  In practice,
we analyze each resimulated halo at the time between redshifts 0.05
and 0 when it is closest to dynamical equilibrium, defined as the time
when the ratio of kinetic to potential energy is closest to 0.5 for
material within the virial radius. [Throughout this paper, we measure
halo masses, $M_{200}$, within a virial radius, $r_{200}$, defined as
the radius of a sphere of mean interior density $200 \,
\rho_{crit}$. Halo circular speeds,
$V_{200}=(GM_{200}/r_{200})^{1/2}$, are also measured at this radius
unless otherwise specified. Numerical experiments show that for
$\Omega=1$ this radius approximately separates the virialized and
infall regions (Cole \& Lacey 1996). For convenience we continue to
use these definitions when $\Omega_0 \ne 1$.]

Once a halo is chosen for resimulation, the particles within its
virial radius are traced back to the initial conditions, where a small
box containing all of them is drawn.  This box is filled with $\sim
32^3$ particles on a cubic grid which are then perturbed using the
waves of the original P$^3$M simulation, together with extra
high-frequency waves added to fill out the power spectrum between the
Nyquist frequencies of the old and new particle grids.  The regions
beyond the ``high-resolution'' box are coarsely sampled with a few
thousand particles of radially increasing mass in order to account for
the large-scale tidal fields present in the original simulation.

This procedure ensures the formation of a clump similar in all respects to
the one selected in the P$^3$M run, except for the improved numerical
resolution. The size of the high-resolution box scales naturally with the
total mass of each system and, as a result, all resimulated halos have
about the same number of particles within the virial radius at $z=0$,
typically between $5,000$ and $10,000$. The extensive tests presented in
Navarro, Frenk \& White (1996) indicate that this number of particles is
adequate to resolve the structure of a halo over approximately two decades
in radius. We therefore choose the gravitational softening, $h_g$, to be $1
\%$ of the virial radius in all cases. (This is the scale-length of a
spline softening; see Navarro \& White 1993 for a definition.) The
tree-code carries out simulations in physical, rather than comoving,
coordinates, and uses individual timesteps for each particle. The
minimum timestep depends on the maximum density resolved in each case,
but was typically $10^{-5} H_0^{-1}$.

As discussed in Navarro, Frenk \& White (1996), numerically convergent
results require that the initial redshift of each run, $z_{init}$, should
be high enough that all resolved scales in the initial box are still in the
linear regime. In order to satisfy this condition, we chose $z_{init}$ so
that the median initial displacement of particles in the high-resolution
box was always less than the mean interparticle separation.  Problems with
this procedure may arise if $z_{init}$ is so high that the gravitational
softening (which is kept fixed in physical coordinates) becomes
significantly larger that the mean initial interparticle separation. We
found this to be a problem only for the smallest masses, $M \lsim
M_{\star}$, in the $n=0$, $\Omega_0=0.1$ model. In this case the initial
redshift prescribed by the median displacement condition is $z_{init} >
700$ and the gravitational softening is then a significant fraction of the
initial box. This can affect the collapse of the earliest progenitors of
these systems and so introduce spurious effects. We therefore limit our
investigation to $M \gsim M_{\star}$ in this particular cosmological model.
Further tests of the effects of particle number, timestep size, and
gravitational softening are given by Tormen et al. (1996). Their results
confirm that the numerical parameters chosen here are adequate to give
stable and accurate results.

\section{Results}

\subsection{Time Evolution}

Figure 1 illustrates the time evolution of halos of different mass
selected from the $\Omega_0=1$, $n=-1$ series. Time runs from top to
bottom and mass increases from left to right. The box size in each
column is chosen so as to contain always approximately the same number
of particles. The redshifts of each snapshot have been chosen so that
the nonlinear mass $M_{\star}$ increases by factors of $4$ from $z_2$
to $z_1$ and from $z_1$ to $z_0=0$. This figure illustrates
convincingly that low-mass halos complete their assembly earlier than
more massive systems.
%Also note that, scaled in this way, the time
%evolution of halos of a given $M/M_{\star}$ is roughly similar
%regardless of the cosmological model.

\subsection{Density Profiles}

Figure 2 shows spherically averaged density profiles at $z=0$ for one
of the least and one of the most massive halos for each set of
cosmological parameters.  These halos span almost four orders of
magnitude in mass in the case of the CDM models, and about two orders
of magnitude in mass in the power-law models. Radial units are kpc for
CDM models (scale at the top), and are arbitrary in the power-law
panels. Density is in units of $10^{10} M_{\odot}/$kpc$^3$ in the CDM
models and in arbitrary units in the others. Solid lines are fits to
each halo profile using eq.~1.  This simple formula provides a good
fit to the structure of all halos over about two decades in radius,
from the gravitational softening (indicated by arrows in Figure 2) to
about the virial radius.  The quality of the fit is essentially
independent of halo mass or cosmological model and implies a
remarkable uniformity in the equilibrium structure of dark matter
halos in different hierarchical clustering models.

The solid and dashed lines in Figure 3 show the profile fits of
Figure~2 but with the radius scaled to the virial radius of each
halo. This scaling removes the intrinsic dependence of size on mass
(more massive halos are bigger) and allows a meaningful comparison
between halos of different mass.  From the definition of virial
radius, the ``concentration'' of a halo, $c=r_{200}/r_s$, and the
characteristic density, $\delta_c$, are linked by the relation
$$
\delta_c={200 \over 3} {c^{3} \over
\bigl[\ln(1+c)-c/(1+c)\bigr]}. \eqno(2)
$$ 
Thus at given halo mass (specified by $M_{200}$), there is a single
free parameter in eq.~1, which may be expressed either as the
characteristic density, $\delta_c$, or as the concentration parameter,
$c$.  This free parameter varies systematically with mass; Figure~3
shows that $c$ and $\delta_c$ decrease with increasing halo mass.

A universal density profile implies a universal circular velocity
profile, $V_c(r)=(GM(r)/r)^{1/2}$.  This is illustrated in Figure~4,
where we plot $V_c$-profiles for the same systems shown in
Figure~2. As in Figure 3, radii are plotted in units of the virial
radius; circular speeds have been normalized to the value at the
virial radius, $V_{200}$. The circular velocity curve implied by eq.~1
is
$$
\biggl({V_c(r)\over V_{200}}\biggr)^2={1 \over x}
{\ln(1+cx)-(cx) /(1+cx) 
\over  \ln(1+c)-c/(1+c)}, \eqno(3)
$$
where $x=r/r_{200}$ is the radius in units of the virial radius.  Circular
velocities rise near the center, reach a maximum ($V_{max}$) at $x_{max}
\sim 2/c$, and decline near the virial radius. More centrally concentrated
halos (higher $\delta_c$ or higher $c$) are characterized by higher values
of $V_{max}/V_{200}$. The dashed lines in Figure~4 show plots of eq.~3
with parameter values derived from the fits to the density profiles of 
Figure~2. The dotted lines are fits using a Hernquist (1990) model
constrained to match the location of the maximum of the $V_c$-curve. The
two fits are indistinguishable near the center, but the Hernquist
model underestimates $V_c$ near the virial radius. This disagreement
becomes more pronounced in lower mass systems, for which 
$\delta_c$ and $V_{max}/V_{200}$ are larger.

\subsection{The mass dependence of halo structure}

The mass-density dependence pointed out above is further illustrated
in Figure~5, where we plot $\delta_c$ versus mass (expressed in units
of $M_{\star}$) for all the systems in each series. An equivalent
plot, illustrating the mass dependence of the concentration, $c$, is
shown in Figure~6. (The panel on the upper-left, corresponding to the
SCDM model, is equivalent to Figure~7 of Navarro, Frenk \& White
1996).  The characteristic density of a halo increases towards lower
masses in all the cosmological models considered. This result supports
the idea that the $M_{200}$-$\delta_c$ relation is a direct result of
the higher redshift of collapse of less massive systems, and suggests
a simple model to describe the mass-density relation. This model
assigns to each halo of mass $M$ (identified at $z=0$) a collapse
redshift, $z_{coll}(M,f)$, defined as the time at which half the mass
of the halo was first contained in progenitors more massive than some
fraction $f$ of the final mass.  With this definition, $z_{coll}$ can
be computed simply using the Press-Schechter formalism (e.g. Lacey \&
Cole 1993),
$$
 {\rm erfc}\biggl({\delta_{crit}(z_{coll}) - \delta_{crit}^0 \over
\sqrt{2(\Delta^2_0(fM)-\Delta^2_0(M))}}\biggr)={1 \over 2}, \eqno(4)
$$
where $\Delta_0^2(M)$ is the linear variance of the power spectrum at
$z=0$ smoothed with a top-hat filter of mass $M$, $\delta_{crit}(z)$
is the density threshold for spherical collapse by redshift $z$, and
$\delta_{crit}^0=\delta_{crit}(0)$. [This definition can be extended
to halos identified at any redshift $z_0$ by replacing
$\delta_{crit}^0$ by $\delta_{crit}(z_0)$ in eq.~4.] Assuming the
characteristic density of a halo to be proportional to the density of
the universe at the corresponding $z_{coll}$ then implies
$$
\delta_c(M | f)= C\,  \Omega_0 \,  \bigl(1+z_{coll}(M,f)\bigr)^3 \eqno(5)
$$
where $C$ is a proportionality constant which might, in principle,
depend on $f$ and on the power spectrum.

We will see below that $f\ll 1$ is needed for this argument to give a
good fit to our simulation data. In this limit $\Delta^2_0(fM) \gg
\Delta^2_0(M)$ and eq.~4 reduces to
$$
\delta_{crit}(z_{coll}) = \delta_{crit}^0 +
C' \Delta_0(fM), \eqno(6)
$$ 
where $C' \approx 0.7$. For $f \ll 1$, $\delta_{crit}(z_{coll}) \gg
\delta_{crit}^0$ for all masses in the range of interest, so that
$\delta_{crit}(z_{coll}) \propto \Delta_0(fM)$. Since
$M_{\star}(z_{coll})$ is defined by
$\Delta_0(M_{\star}(z_{coll}))=\delta_{crit}(z_{coll})$, this equation
implies that the characteristic density of a halo is proportional to
the mean density of the universe at the time when $M_{\star} \approx f
M$, ie. when the characteristic non-linear mass is a fixed small
fraction of the final halo mass. For scale-free models this implies
$\delta_c \propto M^{-(n+3)/2}$, the same scaling that links
$M_{\star}(z)$ and the mean cosmic density at redshift $z$.

Figure 5 shows the correlations predicted from eq.~5 for three values
of the parameter $f$: $0.5$, $0.1$, and $0.01$. The value of the
proportionality constant, $C(f)$, is chosen in each case in order to
match the results of the {\it Einstein-de Sitter} simulations for 
$M=M_{\star}$. These values are given in Table 1. The {\it same} values 
of $C(f)$ are used to plot the curves in the panels
corresponding to the low-density models.  Some interesting results
emerge from inspection of Figure 5 and Table 1.

\items{(i) The agreement between the mass-density dependence predicted by
eq.~5 and the results of the Einstein-de Sitter simulations improves
for smaller values of $f$. This is also true for the low-density
models. Once $C(f)$ is fixed by matching the results of the
Einstein-de Sitter models, the {\it same} value of $C(f)$ provides a
good match to the low density models only if $f \lsim 0.01$.
Interestingly, for $f=0.01$ approximately the same value of the
proportionality constant, $C \approx 3 \times 10^3$, seems to fit all
our simulations.}

\items{(ii)
The characteristic density of $M_{\star}$ halos decreases
systematically for more negative values of the spectral index
$n$. At $M=M_{\star}$, SCDM halos are the least dense in our
$\Omega_0=1$ series, less concentrated still than those corresponding
to $n=-1.5$. This is consistent with the general trend because,
according to eqs. 4 and 5, the characteristic density of a halo of
mass $M_{\star}$ is controlled by the shape of the power spectrum on
scales $\sim f M_{\star}$. This is about $\sim 10^{11} M_{\odot}$ for
$f\approx 0.01$ and the effective slope of the CDM spectrum
on this mass scale is $n_{\rm eff}\sim -2$. }

\items{(iii) For the power-law models with $n=0$ and $-1$ the 
characteristic density at a given $M/M_{\star}$ increases as
$\Omega_0$ decreases. Such a trend is plausible since we expect 
the collapse redshift of halos of a given mass to increase as 
$\Omega_0$ decreases. On the other hand,
halos formed in the low-density CDM$\Lambda$ universe are actually
{\it less} dense than those formed in the standard biased CDM
model because $\delta_c$
depends not only on collapse redshift but also on $\Omega_0$
(see eq.~5). Although reducing $\Omega_0$ increases
the collapse redshift, the increase in $\delta_c$ from the
$(1+z_{coll})^3$ factor can be 
outweighed by the change in $\Omega_0$. In the
CDM$\Lambda$ model the two effects can combine to give a reduction in
$\delta_c$ as $\Omega_0$ decreases. (We remind the reader that 
$\delta_c$ is defined relative to
the critical density rather than the mean density.)}

\items{
(iv) Each halo has a characteristic maximum circular speed, $V_{max}$
(see eq.~3), which is strongly correlated with its mass. This is shown
in Figure 7, where we also plot least squares
fits of the form $M_{200} 
\propto V_{max}^{\alpha}$ to the data in each series.
Consistent with the trends shown in Figures 5 and 6, the correlation
steepens as $n$ increases; we find $\alpha \sim 3.2$ -- $3.3$ for CDM
models and $\alpha > 5$ for $n=0$. Note also that the correlations are
extremely tight; the {\it rms} scatter in $\log M_{200}$ is less than
$\sim 0.1$ in all cases. This is in part a consequence of the
generally good fit of the density profile of eq.~1. The ratio
$V_{max}/V_{200}$ increases only logarithmically with the central
concentration of the halo, and changes only by about a factor of two
as $\delta_c$ varies by four orders of magnitude between $10^3$ and
$10^7$. As a result, the $M_{200}$-$V_{max}$ relation does not deviate
much from the $M_{200}$-$V_{200}$ relation which, by definition, has
zero scatter. This has important consequences for the expected
tightness of empirical correlations between mass and characteristic
velocity, such as the Tully-Fisher relation. We intend to return to
this issue in future work.}

The results in Figures~ 5--7 support our conclusion that the
characteristic density of a halo is controlled mainly by the mean
matter density of the universe at a suitably defined time of collapse.
One important test of this interpretation is to measure $z_{coll}$
directly in the simulations and to compare the result to eq.~5. To do
this we identify clumps at every output time using our
friends-of-friends group-finder with linking length set to $20 \%$ of
the current mean interparticle separation. We then trace the particles
in the most massive clump identified at $z=0$ (which typically has a
mean overdensity of $\sim 200$) and, at each redshift, add up the
total mass in clumps which contain any of these particles and which
are individually more massive than $10 \%$ of the final mass. We
identify $z_{coll}$ with the redshift at which this mass first exceeds
half of the final mass. This is roughly equivalent to the analytic
procedure outlined in eq.~4 for $f=0.1$. (We decided to use $f=0.1$
rather than $f=0.01$ because the smaller value results in very high
collapse redshifts, often before the first output in the simulation.)
The main difference is that some of the mass from the high redshift
progenitors ends up outside the virial radius at $z=0$. This causes a
slight bias of the measured collapse redshifts towards higher values
than the Press-Schechter predictions.

The correlation between $\delta_c/\Omega_0$ and $(1+z_{coll})^3$ 
obtained using this
procedure is shown in Figure 8. All halos in Einstein-de Sitter models are
shown with filled circles; those in open universes with open circles; and
those in the CDM$\Lambda$ model with starred symbols. The solid line is
the relation predicted by eq.~5 for $C=5 \times 10^3$. This is clearly an
excellent approximation to the results of the numerical simulations, and
confirms that the mean matter density of the universe at the time of
collapse is the main factor determining the characteristic density of a
halo.  Note that the value of the proportionality constant is slightly
lower than the values given in Table 1 for $f=0.1$. This difference
compensates for the slight bias towards higher collapse redshifts
introduced by our numerical procedure.

A summary of our results is presented in Figure~9. The panels on the left
compile the fits (for $f=0.01$) to the mass dependence of $\delta_c$
and $c$ in the Einstein-de Sitter models. The panels on the right
compare the Einstein-de Sitter results with those for 
low-density models. As noted above, the typical density of
$M_{\star}$-halos increases with $n$. However, the
difference between models becomes less pronounced at higher masses,
and is almost negligible at $M \, \gsim \, 10 M_{\star}$. Halos of a
given $M/M_{\star}$ in low density universes can have either lower or
higher characteristic densities than their Einstein-de Sitter
counterparts, depending on the competing effects of the collapse
redshift and the value of $\Omega_0$. Note that all the curves in
Figure~9 use the same value, $f=0.01$, and essentially the
same value of the proportionality constant $C \approx 3 \times 10^3$
(see eq.~5). Thus, once calibrated for an Einstein-de Sitter model, it is
possible to apply eqs. 4 and 5 to predict the characteristic density
of halos formed in other hierarchically clustering models. In the 
Appendix we provide a detailed description of how to compute
$\delta_c(M)$ numerically for a variety of cosmologies.

\subsection{The scatter in the correlations}

We examine now the origin of the scatter in the correlations presented
above. In particular, we explore whether {\it at fixed mass} the
dispersion in the measured values of $\delta_c$ is due to variations
either in the collapse redshift or in the global angular momentum of
the system.  As shown by Figures 10 and 11, the bulk of the scatter in
$\delta_c$ at a given $M/M_{\star}$ can be attributed to small
differences in the redshift of collapse.  Figure 10 shows that the
$\delta_c$-deviations from the solid-line fits in Figure 5 (i.e. those
for $f=0.01$) correlate strongly with deviations in the redshift of
collapse. (The latter are measured from least-square fits to the
$M_{200}$ vs $(1+z_{coll})$ correlations measured directly from the
simulations.)  Furthermore, the two residuals seem to correlate just
as expected from eq.~5, i.e. $\Delta \log \delta_c =3 \Delta \log
(1+z_{coll})$. (We note that the magnitude of this scatter may not be
fully representative of the dispersion in halo properties
corresponding to each cosmological model, since the sample has been
selected so as to minimize departures from equilibrium.)  This
relation is indicated by the solid line and is seen to reproduce very
well the trend observed in Figure 10. Figure 11 shows the same
$\delta_c$-residuals of Figure 10, but now as a function of the
scatter in the mass--rotation parameter ($M_{200}$ vs $\lambda$)
correlation. (The rotation parameter is defined by $\lambda=J
|E|^{1/2}/GM^{5/2}$, where $J$ is the angular momentum and $E$ is the
binding energy of the halo. The median $\lambda$ in our series is
$\sim 0.04$, in good agreement with previous studies.) We find no
discernible correlation between $M/M_{\star}$ and $\lambda$ or between
the $\delta_c$- and $\lambda$-residuals (Figure 11). This provides
further evidence supporting our contention that the redshift of
collapse is the primary factor determining the characteristic density
of a halo.

\section{Comparison with previous work}

The main conclusion of our study, that the shape of halo density profiles
is independent of cosmological context, appears to contradict previous
work on this subject. As discussed in \S1, a strong dependence of the slope
of the density profile on the spectral index $n$ and the density parameter
$\Omega_0$ has been established by a number of analytic and numerical
studies. We now show that our results actually include and extend those of
previous workers, and we offer an attractive explanation for some discrepancies
found in the literature.

We first address the claim by Quinn et al. (1986), Efstathiou et
al (1988), Zurek et al. (1988), and Warren et al. (1992) that halo density
profiles steepen with increasing values of the spectral index $n$ in an
Einstein-de Sitter universe.  This claim is based on the discovery
that circular velocity profiles of galaxy-sized halos are relatively
flat for $n \approx -2$ (or for SCDM; Frenk et al. 1985) but decline
progressively faster at large $r$ for larger $n$. Figure 12 shows that
we find the same trend if we analyze our data in the same fashion as
these authors.  In this figure we plot, in linear units, the $V_c$-profile
of an $M_{\star}$ halo for different values of $n$. Each curve has been
computed using eq.~3 and the values of $c$ obtained from the fits presented
in Figure $9$. Since all these halos have the same mass, they also
have the same virial radius and circular speed, $r_{200}$ and $V_{200}$,
respectively, which we have used as normalizing factors. The linear units
in Figure 12 obscure the fact that the form of the curves is the {\it
same} in all cases, and encourage one to conclude, as did previous
authors, that halo rotation curves steepen with increasing $n$. In fact
this trend is present because $M_{\star}$ halos collapse earlier, and
so are {\it denser}, for larger values of $n$.

We note that the trend with $n$ in Figure 12 is sensitive to the choice of
halo mass. Had we used halos with $M < M_{\star}$ the trend would have
been stronger. On the other hand, very massive halos ($M \, \gsim 10
\, M_{\star}$) have similar characteristic densities, irrespective of $n$
(see Figure 9). Thus, had we plotted very massive halos in Figure 12 we
would have concluded that the density profile depends only weakly on
$n$ (at least for $\Omega_0=1$). This is reminiscent of the claim by West
et al. (1987) that the structure of {\it galaxy cluster} halos
is independent of $n$ in Einstein-de Sitter universes. Our
analysis suggests an explanation for this apparently discrepant
result. By considering only the most massive systems in their simulations,
West et al. focused on a mass range where the dependence of $\delta_c$ on
$n$ is minimal, and so concluded (correctly) that cluster
profiles depend very weakly on initial power spectrum.

A similar explanation accounts for the weak dependence
of halo density profile on spectral index $n$ and on $\Omega_0$ reported
by Crone et al. (1994). These authors chose to combine the 35 most massive
clumps in each of their cosmological simulations in order to produce an
``average'' density profile for each value of $n$ and $\Omega_0$.
They then fitted power laws of the form $\rho(r) \propto r^{\gamma}$ 
to these profiles in the radial range corresponding to density contrasts 
between 100 and 3000. For $\Omega_0=1$, this
procedure yielded $\gamma$ values that decrease from $-2.2$ to $-2.5$ as $n$
increases from $-2$ to $0$ (see the curve labeled ``EdS'' in their
Figure 4). Our results show this weak trend to be a consequence of the
averaging procedure they adopted. The shape of the halo mass function
depends strongly on $n$ and is increasingly peaked around
$M \approx M_{\star}$ for larger values of $n$. Thus, combining the 35 most
massive clumps in a simulation results in different ``effective'' masses
for different values of $n$, with a bias towards larger values of
$M/M_{\star}$ for more negative values of $n$. Applying the same selection
procedure as Crone et al. to our own cosmological simulations we find that the
median mass of these halo ensembles increases from $\sim M_{\star}$ for
$n=0$ to $\sim \, 6 M_{\star}$ for $n=-1.5$. As illustrated in Figure 13,
fitting power-laws to the density profiles of these ``average'' halos
results in a weak steepening with $n$ similar to the trend reported by
Crone et al. The slopes of $\gamma\sim -3$ which they found for low
density models are also easily understood, since in these cases they
fit a radial range which extends well outside our nominal virial
radius $r_{200}$.

Our results are also in agreement with recent work by Cole \& Lacey
(1996) and Tormen et al. (1996), who analyzed the structure of massive
halos ($M>M_{\star}$) in scale-free universes. The general
correlations with mass which they report (more massive halos tend to be less
centrally concentrated than less massive halos) agree well with the 
results we present in \S
3. At a given halo mass, however, Cole \& Lacey's halos are
significantly less concentrated than ours. For example, for 
$M_{\star}$ halos they find $c \sim 12$ and $\sim 17$ for $n=-1$ and $0$,
respectively ($\Omega_0=1$), compared with $c \sim 17$ and $\sim 30$
in our simulations (see connected symbols in the lower-right panel of
Figure 9).  On the other hand, the results of Tormen et al. (1996) are
in very good agreement with ours; they find $c \sim 10$ for a $10
M_{\star}$ halo formed in an $n=-1$, $\Omega_0=1$ universe, which
compares well with $c \sim 9$ that we find for a similar system.

As discussed in detail by Tormen et al, the discrepancy between our
results and those of Cole \& Lacey is likely to be due in part to the
poorer numerical resolution of their simulations. Indeed, their
simulations used gravitational softenings that are $2$-$3$ times larger
than ours and timesteps which are also significantly longer than the
typical values in our simulations. Both of these effects can
artificially lower the central concentration of a halo. A
concurrent factor may be the averaging procedure adopted by Cole \&
Lacey. These authors constructed {\it average} density profiles by
co-adding all halos of similar mass identified in their cosmological
simulations, regardless of their dynamical state. This sample 
contains a number of unrelaxed systems and ongoing mergers where
substructure and double centers can bias the results of the fitting
procedure towards lower concentrations.

We can test directly for the effects of these various factors by
applying the averaging procedure of Cole \& Lacey to halos identified
in our own $P^3M$ cosmological simulations (see \S2.2.1) and comparing
with the results of the individual halo runs. We restrict this
comparison to the most massive halos in each run, $M \, \gsim \, 10
M_{\star}$, since these systems have a large number of particles and
comparable numerical resolution to that of Cole \& Lacey. The
comparison confirms that the central concentration of halos can be
significantly underestimated as a result of the factors mentioned
above. The magnitude of the effect is sensitive to $n$ and $\Omega_0$;
the concentration $c$ can be underestimated by up to a factor of $\sim
3$ for $n=0$ and $\Omega_0=0.1$, but by less than $\sim 1.5$ for
$n=-1$ and $\Omega_0=1$. We conclude that the disagreement between our
results and those of Cole \& Lacey is likely to be the result of the
combined effect of their halo selection and averaging procedures and
their poorer numerical resolution.

\section{Discussion}

Our results suggest that equilibrium dark halos formed through
dissipationless hierarchical clustering have density profiles with a
universal shape which does not depend on their mass, on the power
spectrum of initial fluctuations, or on the cosmological parameters
$\Omega_0$ and $\Lambda$.  It appears that mergers and collisions act
during halo formation as a ``relaxation'' mechanism to produce an
equilibrium which is largely independent of initial conditions.  This
mechanism must operate rapidly since the similarity between profiles
extends to the virial radius. These properties are characteristic of
the ``violent relaxation'' process proposed by Lynden-Bell (1967) to
explain the regularities observed in the structure of elliptical
galaxies, and it is interesting that our universal profile is similar
to the Hernquist profile which gives a good description of elliptical
galaxy photometry. The two differ significantly only at large radii,
perhaps because ellipticals are relatively isolated systems whereas
dark halos are not. Syer \& White (1996) suggest that a universal
profile may be understood as a fixed point for a process of repeated
mergers between unequal objects. Their analysis of this process
predicts a dependence on initial power spectrum which seems stronger,
however, than that seen in our numerical data.
%The mass profile of halos is thus analogous to
%other properties of virialized systems formed by hierarchical
%clustering such as the spins and the three-dimensional shapes, whose
%distributions are essentially independent of initial conditions
%although their individual values depend (albeit weakly) on mass and
%cosmological parameters (see Cole \& Lacey 1996 and references
%therein).

Our simulations suggest that the density profile of an isolated
equilibrium halo can be specified quite accurately by giving two
parameters; halo mass and halo characteristic density. Furthermore, 
in any particular hierarchical model these parameters are related
in such a way that the characteristic density is
proportional to the mean cosmic density at the
time when the mass of typical nonlinear objects was some fixed small
fraction of the halo mass. The characteristic density thus
reflects the density of the universe at the collapse time of the 
objects which merge to form the halo core. With this interpretation we
are able to fit the mass-density relation for equilibrium halos in
all the cosmological models we have considered. In addition, it is
possible to calculate the mass-density relation in any other hierarchical
model (see Appendix).  It is difficult to imagine a simpler situation
-- halos of all masses in all hierarchical cosmologies look the same,
and their characteristic densities are just proportional to the cosmic
density at the time they ``formed''.

Our results extend the radial range over which dark halo structure can
reliably be determined to more than two decades. The central regions
of our models have densities of order $10^6 \rho_{crit}$, comparable
to those in the luminous parts of galaxies and in the central cores of
galaxy clusters. As a result a variety of direct observational tests
of our predictions are available.  In Navarro et al. (1996) we
discussed several of these in the context of the standard CDM
cosmogony -- rotation curves of giant and dwarf galaxies, satellite
galaxy dynamics, hot gaseous atmospheres around galaxies and in
clusters, strong and weak gravitational lensing -- all provide
interesting constraints. We will not pursue these issues here,
however, because they would take us too far from our primary goal,
namely the presentation of a simple and apparently general theoretical
result: hierarchical clustering leads to a universal halo density
profile just as it leads to universal distributions of halo axial
ratios and halo spins; none of these properties depends strongly on
power spectrum, on $\Omega$, or on $\Lambda$ [see Cole \& Lacey (1996)
and references therein]. Comparison of the predicted halo structure
with observation should, of course, provide strong constraints on the
parameters which define particular hierarchical cosmogonies, and
perhaps on the hierarchical clustering paradigm itself. We expect to
come back to these issues in future work.

\acknowledgments
{We are grateful for the hospitality of the Institute for Theoretical
Physics of the University of California at Santa Barbara, where some
of the work presented here was carried out. JFN is also grateful for
the hospitality of the Max Planck Institut f\"ur Astrophysik in
Garching, where this project was started. In addition he would like to
acknowledge useful discussions with Cedric Lacey. Special thanks are
due to Shaun Cole for making the data shown in Figure 9 available in
electronic form. This work was supported in part by the PPARC and by
the NSF under grant No. PHY94-07194 to the Institute for Theoretical
Physics of the University of California at Santa Barbara.}

\appendix
\section{
Step-by-step calculation of the density profile of a dark matter halo}

In this Appendix we describe in detail the calculation of the
parameters that specify the density profile of a dark halo of mass
$M$. This calculation is applicable to Einstein-de Sitter ($\Omega_0=1$,
$\Lambda=0$), open ($\Omega_0 < 1$, $\Lambda=0$), and flat
($\Omega_0+\Lambda=1$) universes. We provide approximate fitting
formulae that are valid for power-law and CDM initial density
fluctuation spectra.

A halo of mass $M$ identified at $z=z_0$ can be characterized by its
virial radius,
$$
r_{200}=1.63 \times 10^{-2} \biggl({M\over h^{-1}
M_{\odot}}\biggr)^{1/3} 
\biggl({\Omega_0 \over \Omega(z_0)} \biggr)^{-1/3}
(1+z_0)^{-1} \,
h^{-1}{\rm kpc}, \eqno(A1)
$$
or by its circular velocity, 
$$
V_{200}=\biggl({G M \over r_{200}}\biggr)^{1/2} =
\biggl({r_{200} \over h^{-1} {\rm kpc}}\biggr) \, 
\biggl({\Omega_0 \over \Omega(z_0)}\biggr)^{1/2} (1+z_0)^{3/2} \,
 {\rm km/s}. \eqno(A2)
$$
The density profile of this system is fully specified by its
characteristic density $\delta_c$ and is given by (see eq.1)
$$
\rho(r)={3 H_0^2 \over 8 \pi G} \,
(1+z_0)^3 \, {\Omega_0 \over \Omega(z_0)} \, {\delta_c \over
 cx(1+cx)^2}, \eqno(A3) $$
where $x=r/r_{200}$ and $c$ is the concentration parameter, a function
of $\delta_c$ given in eq.~2. The corresponding circular velocity
profile, $V_c(r)$, is given by
$$
\biggl({V_c(r)\over V_{200}}\biggr)^2={1 \over x}
{\ln(1+cx)-(cx) /(1+cx) 
\over  \ln(1+c)-c/(1+c)}, \eqno(A4)
$$
using the concentration $c$ as a parameter.

The characteristic density is determined by the collapse redshift
$z_{coll}$, which is given by (see eq.~4)
$$ {\delta_{crit}(z_{coll}) \over \delta_{crit}(z_0)} =
{\delta_{crit}^0(\Omega(z_{coll}),\Lambda) 
\over \delta_{crit}^0(\Omega(z_0),\Lambda)}
{D(z_0,\Omega_0,\Lambda) 
\over D(z_{coll},\Omega_0,\Lambda)} = 
1 + {0.477 \over \delta_{crit}(z_0)}
\sqrt{2(\Delta_0^2(fM) -
\Delta_0^2(M))} \eqno(A5) 
$$
%
%The critical density threshold for collapse, $\delta_{crit}$, is given by
%%
%$$
%\delta_{crit}(z|\Omega_0,\Lambda)=
%\cases{
%{\delta_{crit}^0/D(z)}     
%                            & if $\Omega_0=1$ and $\Lambda=0$,\cr
%{(3/2)D(0)(1+(t_{\Omega}/t(z))^{2/3})}
%                            &if $\Omega_0<1$ and $\Lambda=0$,\cr
%{\delta_{crit}^0(\Omega(z))/D(z)}
%                            & if $\Omega_0+\Lambda=1$,\cr} \eqno(A6)
%$$
%%
%Here $t_{\Omega}=\pi H_0^{-1} \Omega_0 (1-\Omega_0)^{-3/2}$ and
Here $D(z,\Omega_0,\Lambda)$ is an $\Omega$-dependent linear growth factor
that can be written as
$$
D(z,\Omega_0,\Lambda)=
\cases{
{1/(1+z)}                         & if $\Omega_0=1$ and $\Lambda=0$,\cr
{F_1(w)/F_1(w_0)}                 & if $\Omega_0<1$ and $\Lambda=0$,\cr
{F_2(y) F_3(y)/F_2(y_0) F_3(y_0)} & if $\Omega_0+\Lambda=1$,\cr} \eqno(A6)
$$
where we have used the following auxiliary definitions,
$$
w_0={1 \over \Omega_0} - 1, \eqno(A7)
$$
$$
w={w_0 \over 1+z}, \eqno(A8)
$$
$$ F_1(u)=1+{3 \over u} + {3(1+u)^{1/2} \over u^{3/2}}
\ln{\bigl[(1+u)^{1/2} -u^{1/2}\bigr]}, \eqno(A9) 
$$
$$
y_0=(2 w_0)^{1/3}, \eqno(A10)
$$
$$
y={y_0 \over 1+z}, \eqno(A11)
$$
$$
F_2(u)={(u^{3}+2)^{1/2} \over u^{3/2}}, \eqno(A12)
$$
and
$$
F_3(u)=\int_0^u \biggl({u' \over u'^3+2}\biggr)^{3/2} du'. \eqno(A13)
$$
A good numerical approximation to the $\Omega$-dependence of the
critical threshold for spherical collapse is given by
$$
\delta_{crit}^0 (\Omega,\Lambda)=
\cases{
0.15\, (12 \pi)^{2/3}   & if $\Omega=1$ and $\Lambda=0$,\cr
0.15\, (12 \pi)^{2/3}\, \Omega^{0.0185}& if $\Omega_0<1$ and $\Lambda=0$,\cr
0.15\, (12 \pi)^{2/3}\, \Omega^{0.0055}& if $\Omega_0+\Lambda=1$,\cr} \eqno(A14)
$$
which can be used to compute
$\delta_{crit}(z_0)=\delta_{crit}^0(\Omega(z_0))/D(z_0,\Omega_0,\Lambda)$.
Finally, eq.~A5 requires $\Delta_0^2(M)$, the variance of the power
spectrum on mass scale $M$, extrapolated linearly to $z=0$. In the
case of a power-law spectrum of initial density fluctuations, $P(k)
\propto k^n$, this is simply
$$
\Delta_0^2(M)=
\delta_{crit}^0\biggl({M \over M_{\star}(z=0)}\biggr)^{-(n+3)/6}, \eqno(A15)
$$
where we have normalized the spectrum by $M_{\star}(z=0)$, the present
nonlinear mass. A CDM spectrum is usually normalized by $\sigma_8$,
the rms mass fluctuations within a sphere of radius $8 h^{-1}$ Mpc,
and its variance can be approximated by,
$$
\Delta_0(M)=\sigma_8 F_4(M_8)/F_4(M_h), \eqno(A16)
$$
where we have used the following definitions,
$$
M_8=6.005 \times 10^{14} (h \, \Omega_0)^3, \eqno(A17)
$$
$$
M_h=\biggl({M\over h^{-1} M_{\odot}}\biggr) h^3 \Omega_0^2 \eqno(A18)
$$
and
$$
F_4(u)=
A_1 u^{0.67} \bigl[ 1+ (A_2 u^{-0.1} + A_3 u^{-0.63})^{p}\bigr]^{1/p}, \eqno(A19)
$$
with $A_1=8.6594 \times 10^{-12}$, $A_2=3.5$, $A_3=1.628 \times 10^9$,
and $p=0.255$.

Eqs.~A6-A19 can be used to solve eq.~A5 and find the collapse redshift,
$z_{coll}$, corresponding to a halo of mass $M$. As noted when
discussing Fig.~5, we recommend using $f=0.01$ when solving eq.~A5,
since this value seems to reproduce well the results of all our
numerical experiments.

Once $z_{coll}(M)$ has been found, the characteristic density of the
halo (expressed in units of the critical density at $z=z_0$) can be
computed from (see eq.5 and Table 1),
$$
\delta_c(M,z_0) \sim 3 \times 10^3 \, \Omega(z_0) \biggl({1+z_{coll} \over 1+z_0}\biggr)^3, \eqno(A20)
$$
where we have assumed $f=0.01$. 
A FORTRAN subroutine that implements the procedure described here and
returns $\delta_c(M,z_0)$ for all the cosmologies we discuss is
available from the authors upon request.
%
%
%
%
% That's the end of the main body of the paper.  Now we will have some
% back matter.
%
% Tables are usually supposed to be submitted one per page, following
% the main body of the text, so before each table we would have a
% \clearpage to force a page break at that point.  There should also
% be a \clearpage after the last table so that it gets forced onto
% its own page, too.
%
% The tabular data is aligned within the "tabular" environment.  Observe
% that our tabular environment is embedded within a "center" environment,
% which is in turn inside a "table" environment.  Exercise for the reader:
% Why do you think we used the "table*" environment?
%
% We need the table environment for autonumbering and caption generation,
% which is why it is not enough to have a centered tabular.
%
% Within the tabular environment, please note that we use no vertical
% rules, and the only horizontal rule is the \tableline (*not* an \hline)
% which delimits the column headings from the tabular data.  Also note
% that a couple of the column headings require special annotation, i.e.,
% footnotes for tables.  They are marked and tagged with \tablenotemark.
% \tablenotemarks could be placed on individual data entries as well,
% but be careful not to go berserk doing this.

\clearpage

%\include{../../macro/aas/samp2tbl}

% This is the last table for this paper (as well as the first), so we
% should follow it with a \clearpage.  In order to force all the floating
% tables out of their buffers and onto vertical page lists, we must use
% \clearpage rather than \newpage.

\begin{table*}
\begin{center}
\begin{tabular}{crrrrr}
$P(k)$ & $\Omega_0$ & $\Lambda$ & $f$ & $C(f)$ &\\
\tableline
CDM          & 1.0  & 0.0  & 0.5  & $1.75 \times 10^4$ & \\
             & 1.0  & 0.0  & 0.1  & $7.44 \times 10^3$ & \\
             & 1.0  & 0.0  & 0.01 & $3.41 \times 10^3$ & \\
CDM$\Lambda$ & 0.25 & 0.75 & 0.5  & $1.75 \times 10^4$ & \\
             & 0.25 & 0.75 & 0.1  & $7.44 \times 10^3$ & \\
             & 0.25 & 0.75 & 0.01 & $3.41 \times 10^3$ & \\
\\
$n=-1.5$  & 1.0 & 0.0 & 0.5  & $3.00 \times 10^4$ & \\
          & 1.0 & 0.0 & 0.1  & $1.08 \times 10^4$ & \\
          & 1.0 & 0.0 & 0.01 & $3.15 \times 10^3$ & \\
\\
$n=-1.0$  & 1.0 & 0.0 & 0.5  & $5.00 \times 10^4$ & \\
          & 1.0 & 0.0 & 0.1  & $1.38 \times 10^4$ & \\
          & 1.0 & 0.0 & 0.01 & $2.50 \times 10^3$ & \\
          & 0.1 & 0.0 & 0.5  & $5.00 \times 10^4$ & \\
          & 0.1 & 0.0 & 0.1  & $1.38 \times 10^4$ & \\
          & 0.1 & 0.0 & 0.01 & $2.50 \times 10^3$ & \\
\\
$n=-0.5$  & 1.0 & 0.0 & 0.5  & $1.25 \times 10^5$ & \\
          & 1.0 & 0.0 & 0.1  & $2.81 \times 10^4$ & \\
          & 1.0 & 0.0 & 0.01 & $2.81 \times 10^3$ & \\
\\
$n=0.0$   & 1.0 & 0.0 & 0.5  & $4.00 \times 10^5$ & \\
          & 1.0 & 0.0 & 0.1  & $6.66 \times 10^4$ & \\
          & 1.0 & 0.0 & 0.01 & $4.00 \times 10^3$ & \\
          & 0.1 & 0.0 & 0.5  & $4.00 \times 10^5$ & \\
          & 0.1 & 0.0 & 0.1  & $6.66 \times 10^4$ & \\
          & 0.1 & 0.0 & 0.01 & $4.00 \times 10^3$ & \\
\end{tabular}
\end{center}
%
% Text for table footnotes must follow the tabular environment but must
% be inside the table environment.  Note that it is OK to put \ref's
% in \tablenotetext's.
%
%\tablenotetext{a}{Sample footnote for table~\ref{tbl-1}}
%\tablenotetext{b}{Another sample footnote for table~\ref{tbl-1}}
%\tablenotetext{c}{Footnote with no call out}
%\tablenotetext{d}{Another footnote with no call out}
%\tablenotetext{e}{A further additional footnote with no call out}

% The caption contains only the caption text.  The "Table N." identification
% is generated by the \caption command on its own.
%
% It is necessary to \label tables and figures *after* the \caption has been
% specified because the table/figure number is generated by \caption, not
% by \begin{whatever}.

\caption{Parameters in eq.~5 used to plot the fits in Figure 5.} \label{tbl-1}

\end{table*}

\clearpage

% Now comes the reference list.  In this document, we used \cite to call
% out citations, so we must use \bibitem in the reference list, which
% means we use the LaTeX thebibliography environment.  Please note that
% \begin{thebibliography} is followed by a null argument.  If you forget
% this, mayhem ensues, and LaTeX will say "Perhaps a missing item?" when
% you run it.  Do not call us, do not send mail when this happens.  Put
% the silly {} after the \begin{thebibliography}.
%
% Each reference has a \bibitem command to define the citation format
% and the symbolic tag, as well as a \reference command which sets up
% formatting parameters for the reference list itself.
%
% If we had not bothered with the \cite-\bibitem business, calling out
% the references ourselves, the reference list could be enclosed in
% a references environment (\begin{references} has no null argument),
% and there would be no need for the leading \bibitem's.

% Finally, we have figure captions.  Usually these must be on a separate
% page, although unlike table, it is often permissible to have several
% figure captions on the same page.  We force the page break between
% the reference list and the figure captions.
%
% The \caption command in the figure environment works like the one in the
% table environment (it's the same one, actually), except that this one
% produces identification text that reads "Figure N."

\clearpage

\begin{figure}
%\plotone{scfg-1eqn_xy.ps}
\plotone{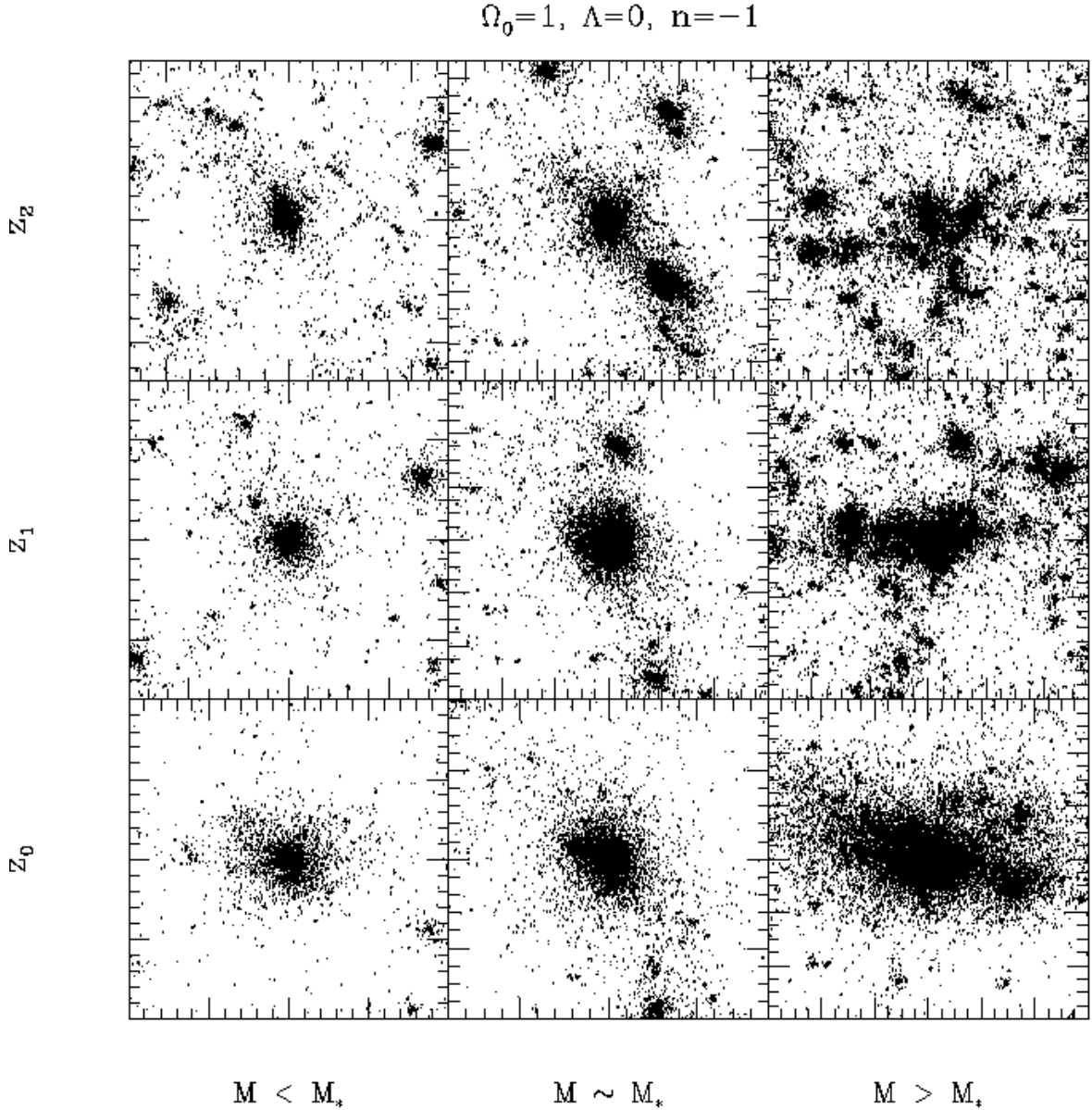}
\figurenum{1}
\caption{
Particle plots illustrating the time evolution of halos of different
mass in an $\Omega_0=1$, $n=-1$ cosmology. Box sizes of each column
are chosen so as to include approximately the same number of
particles. At $z_0=0$ the box size corresponds to about $6
\times r_{200}$. Time runs from top to bottom. Each
snapshot is chosen so that $M_{\star}$ increases by a factor of $4$
between each row. Low mass halos assemble earlier than their more
massive counterparts. This is true for every cosmological scenario in
our series.}
\end{figure}

%\begin{figure}
%\plotone{scfg-1o_xy.ps}
%\figurenum{1b}
%\caption{same as Figure 1a.}
%\end{figure}

%\begin{figure}
%\plotone{cd7_xy.ps}
%\figurenum{1c}
%\caption{same as Figure 1a.}
%\end{figure}

%\begin{figure}
%\plotone{cdml_xy.ps}
%\figurenum{1d}
%\caption{same as Figure 1a.}
%\end{figure}

\begin{figure}
%\plotone{rhoprof.ps}
\plotone{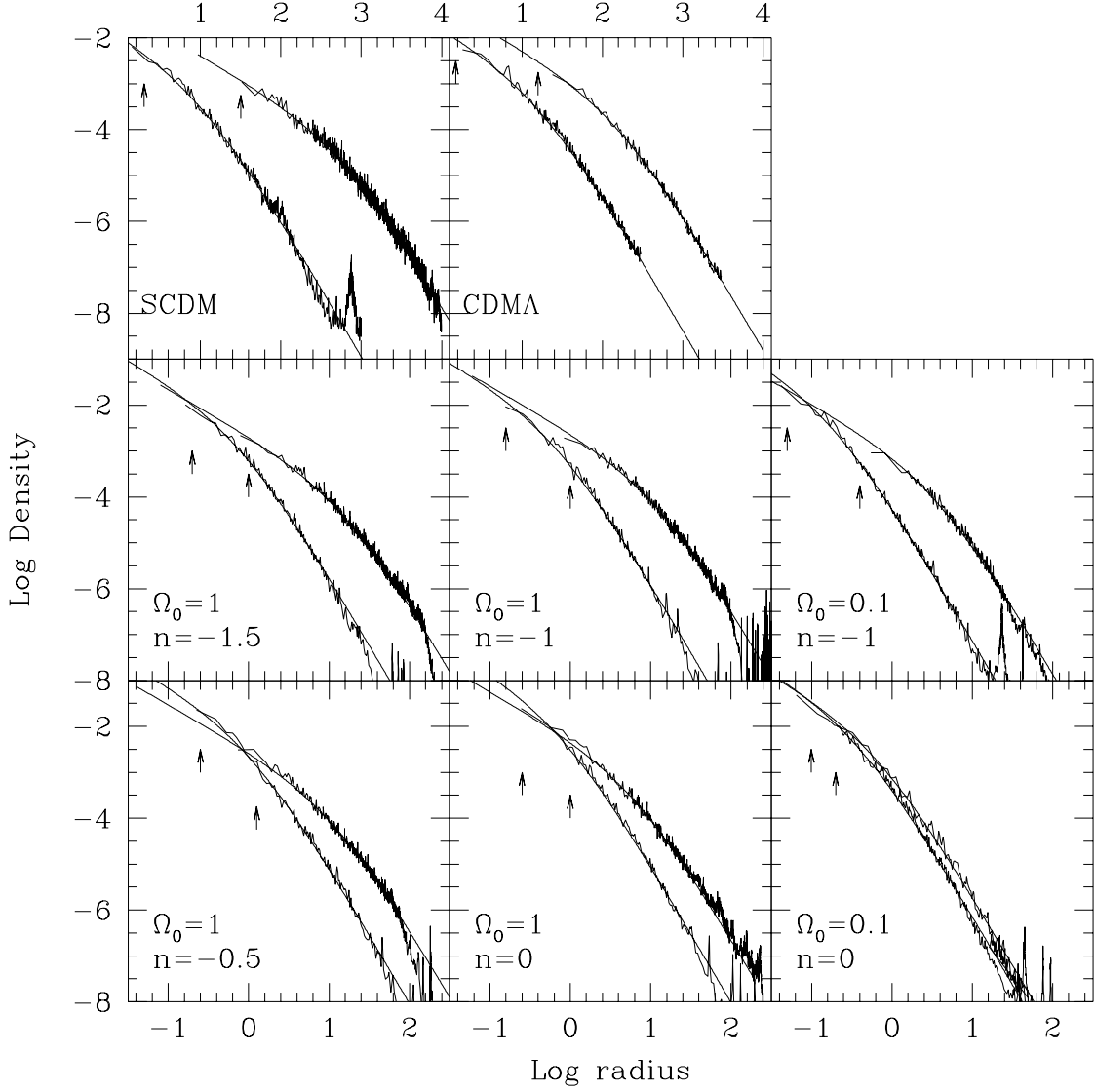}
\figurenum{2}
\caption{
Density profiles of one of the most and one of the least massive halos
in each series. In each panel the low-mass system is represented by
the leftmost curve.  In the SCDM and CDM$\Lambda$ models radii are
given in kpc (scale at the top) and densities are in units of $10^{10}
M_{\odot}$/kpc$^3$. In all other panels units are arbitrary. The
density parameter, $\Omega_0$, and the value of the spectral index,
$n$ is given in each panel. Solid lines are fits to the density
profiles using eq.~(1). The arrows indicate the value of the
gravitational softening. The virial radius of each system is in all
cases two orders of magnitude larger than the gravitational
softening.}
\end{figure}

\begin{figure}
%\plotone{rhoprof_sc.ps}
\plotone{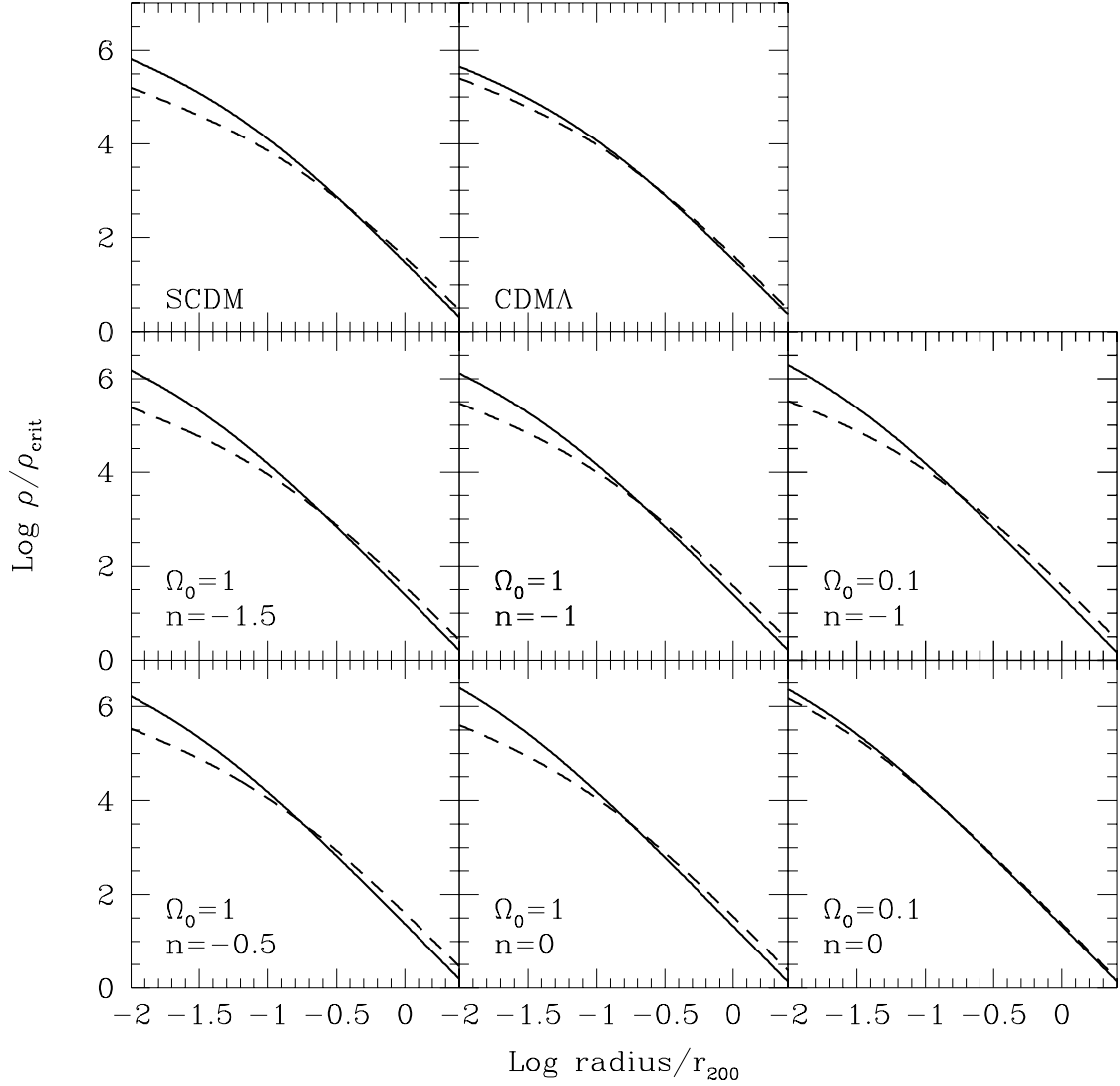}
\figurenum{3}
\caption{
The fits to the density profiles of Figure 2, scaled to the
virial radius, $r_{200}$, of each system and to the critical density
of the universe at $z=0$. Solid and dashed lines correspond to the
low- and high-mass systems, respectively. Note that low-mass systems
are denser than high-mass systems near the center, indicating that the
characteristic density of a halo increases as the halo mass decreases.} 
\end{figure}

\begin{figure}
%\plotone{vcprof_sc.ps}
\plotone{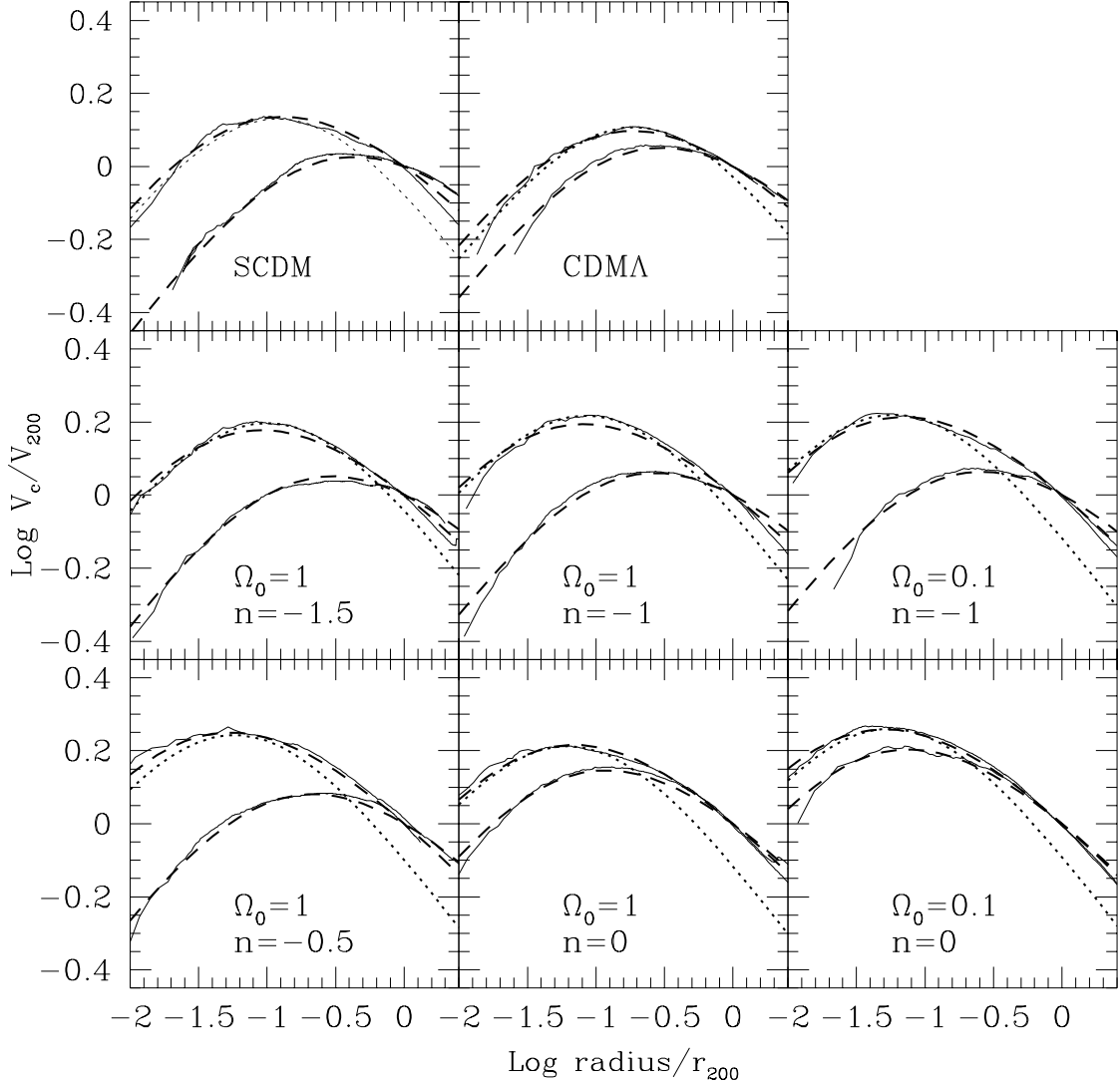}
\figurenum{4}
\caption{
The circular velocity profiles of the halos shown in Figure 2. Radii
are in units of the virial radius and circular speeds are
normalized to the value at the virial radius. The thin solid line
shows the data from the simulations. All curves have the same shape: 
they rise near the center until they reach a maximum and then decline at the
outer edge. Low mass systems have higher maximum circular velocities
in these scaled units because of their higher central
concentrations. Dashed lines are fits using eq.(3). The dotted lines are
the fit to the low-mass halo in each panel using a Hernquist
profile. Note that this model fits rather well the inner regions of
the halos, but underestimates the circular velocity near the
virial radius.}
\end{figure}

\begin{figure}
%\plotone{m200rhoc.ps}
\plotone{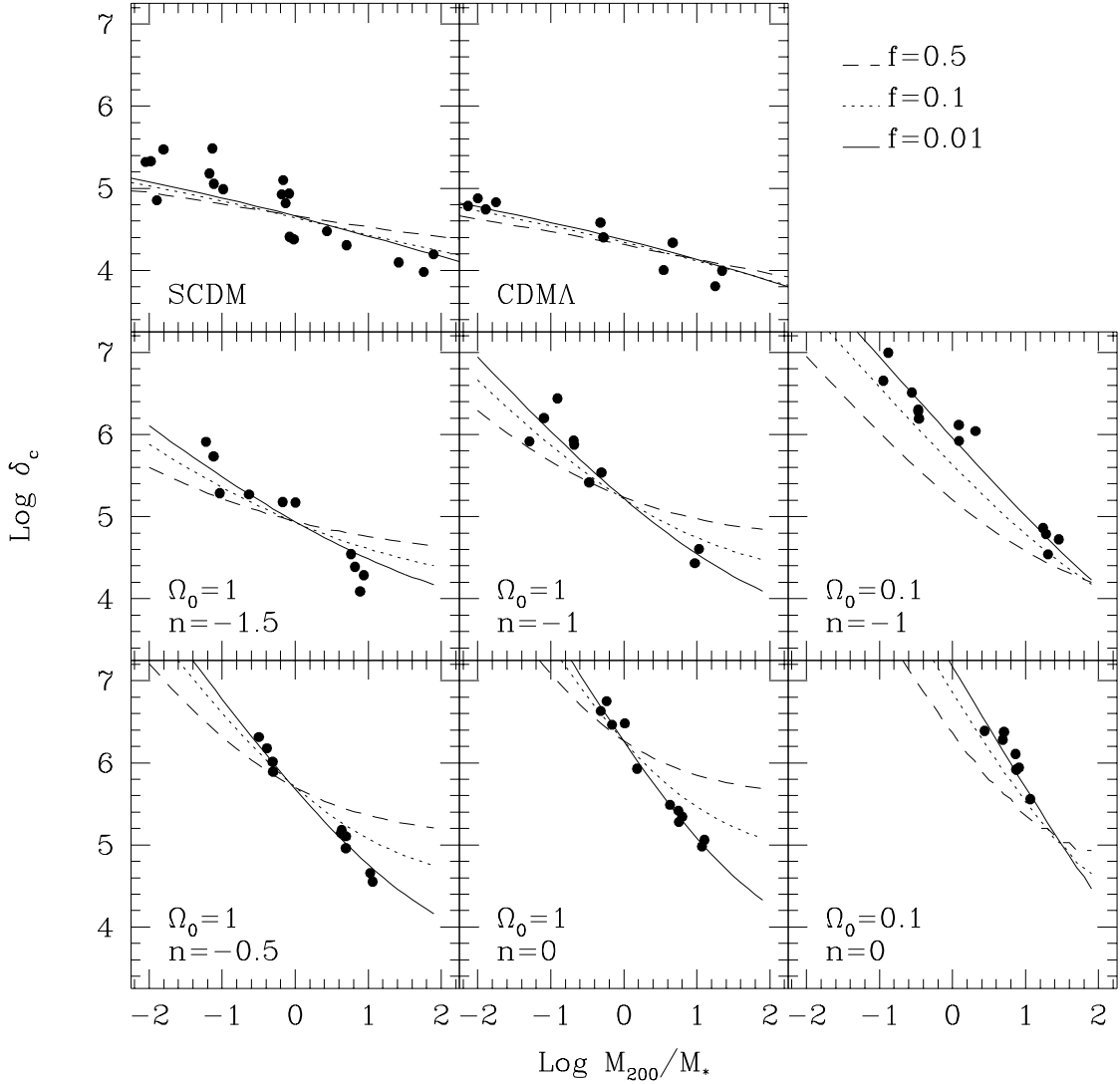}
\figurenum{5}
\caption{
The correlation between the mass of a halo and its characteristic
density. Masses are given in units of the nonlinear mass scale
$M_{\star}$ (see text for a definition). Densities are relative to
the critical value. Three curves are shown in each panel for different
values of the parameter $f$ (see eq.~5). The fits are normalized to
intersect at $M_{200}=M_{*}$ in the case $\Omega=1$. This
normalization is then used for the low-density models ($\Omega_0
<1$). Note that for $f=0.01$ this procedure results in good fits to
the results of the simulations in all cases. }
\end{figure}

\begin{figure}
%\plotone{m200c.ps}
\plotone{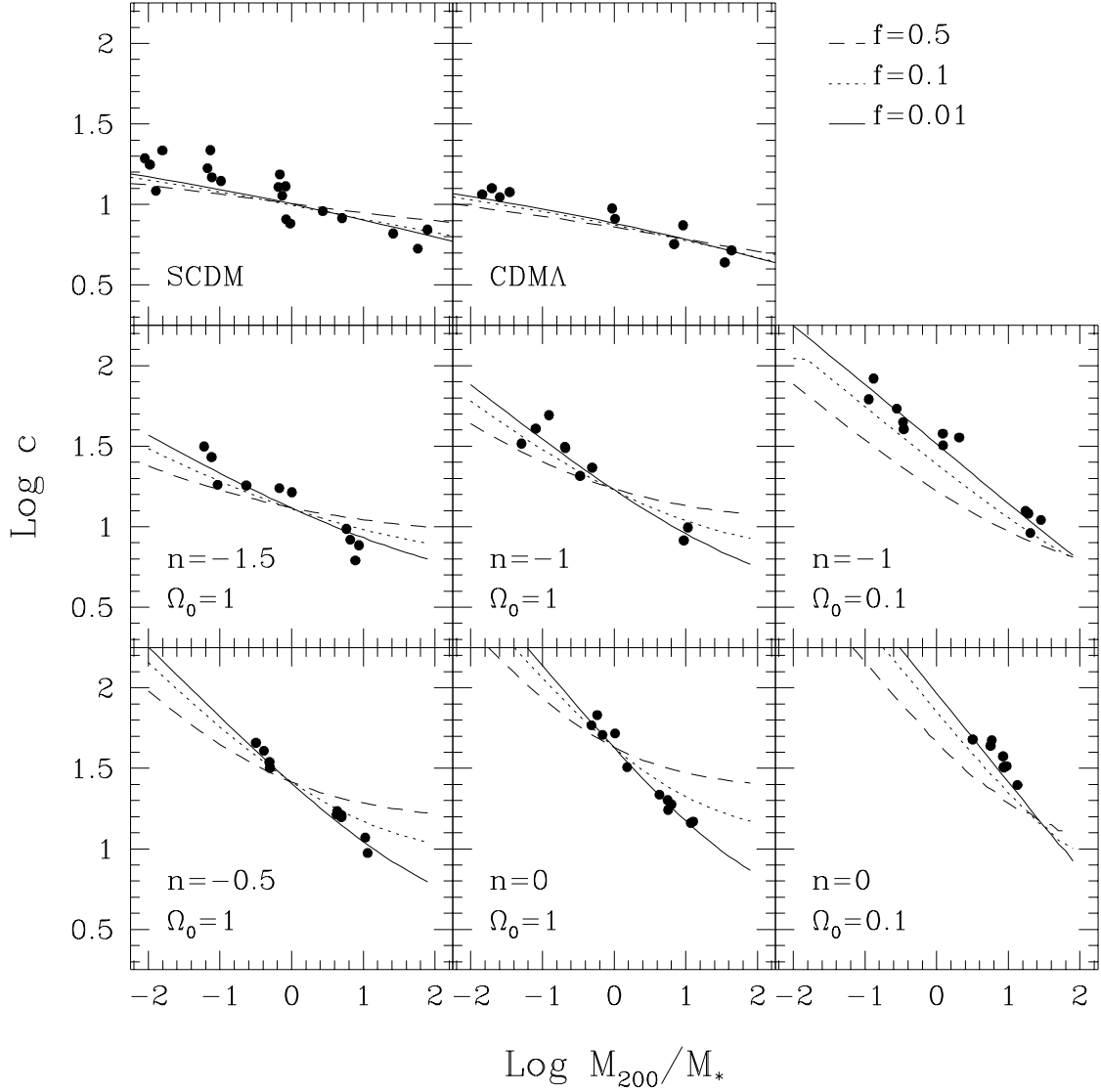}
\figurenum{6}
\caption{
As Figure 5, but for the concentration parameter $c$.  }
\end{figure}

\begin{figure}
%\plotone{m200vmax.ps}
\plotone{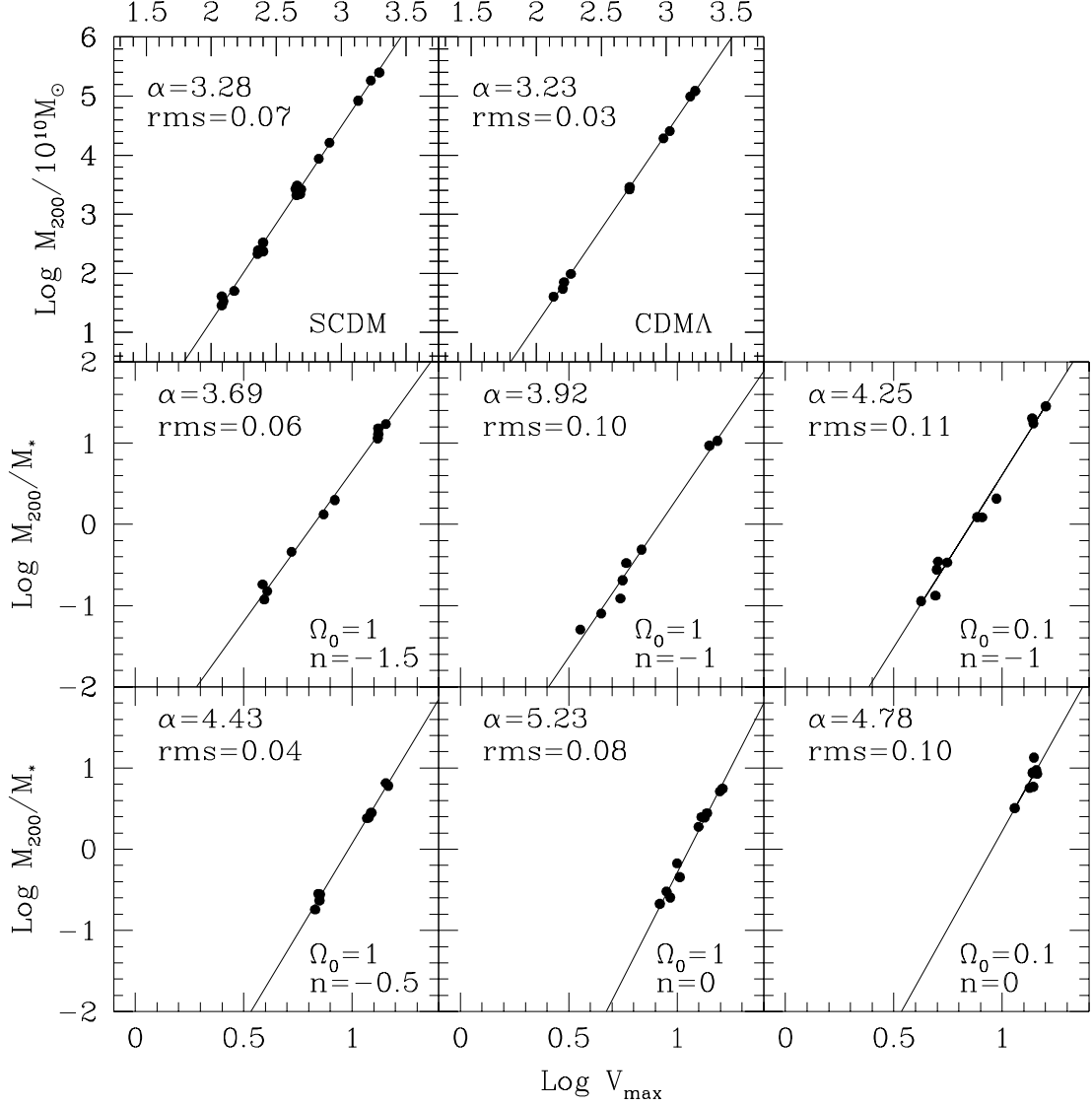}
\figurenum{7}
\caption{
The mass dependence of the maximum circular velocity of a halo.
Velocity units are arbitrary in the power-law panels and km s$^{-1}$
in the CDM models (scale at top). Mass is in units of $10^{10}
M_{\odot}$ for CDM halos and in units of $M_{\star}$ in the other
panels.  Power law fits of the form $M \propto V_{max}^{\alpha}$ are
shown. The value of $\alpha$ and the rms scatter in the mass about the 
fit are indicated in each panel. Note that the $M$-$V_{max}$
dependence steepens for larger values of the spectral index $n$. The
effect of the cosmological parameters on $\alpha$ seems to be 
rather small.}
\end{figure}

\begin{figure}
%\plotone{deltac_zcoll_sim.ps}
\plotone{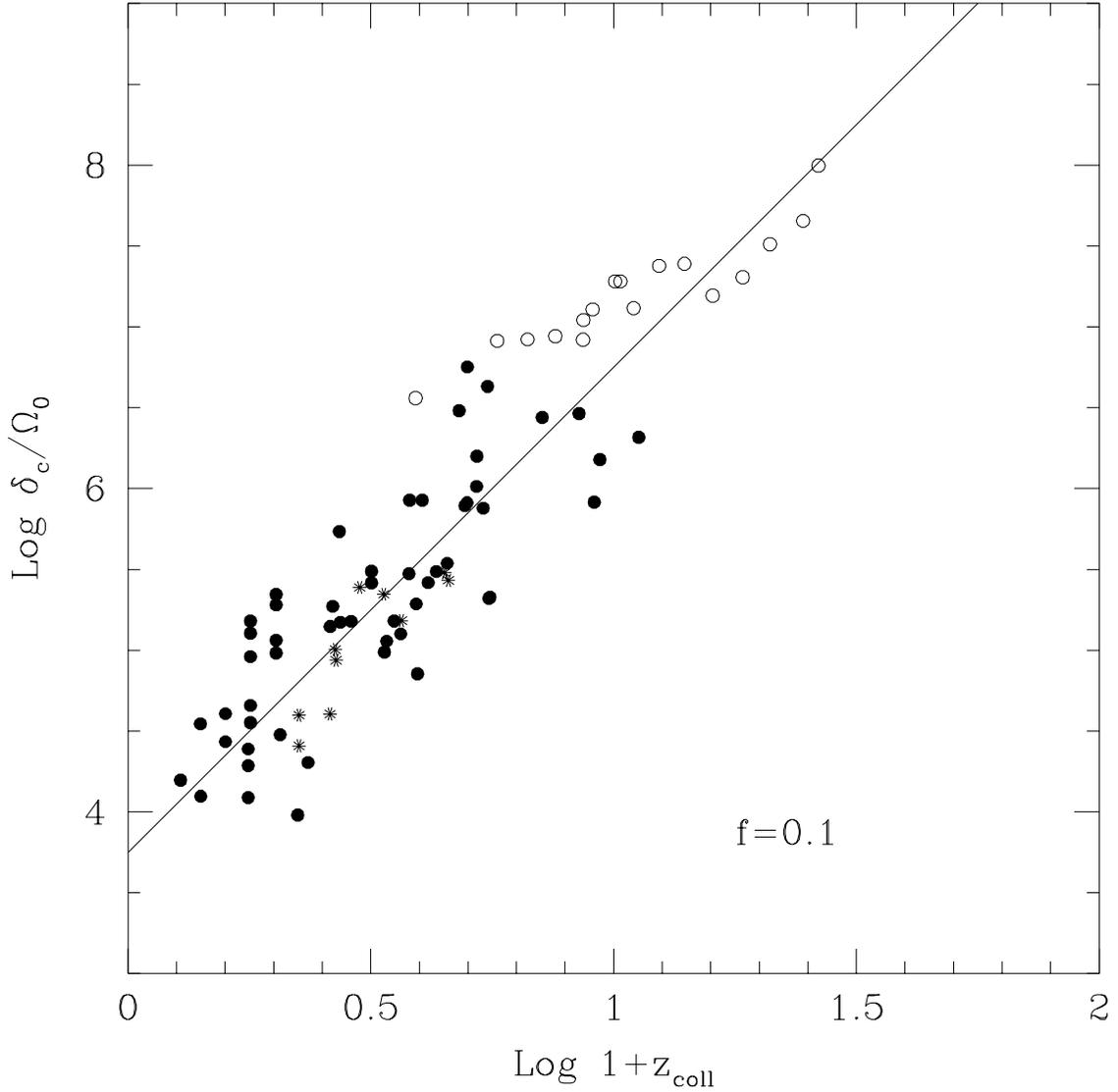}
\figurenum{8}
\caption{
The characteristic density of all halos in our series as a function of
the redshift at which half of the final mass is in collapsed
progenitors more massive than $10 \%$ of the final mass. Solid circles
correspond to all our runs with $\Omega_0=1$, open circles to our runs
with $\Omega_0=0.1$ and $\Lambda=0$, and starred symbols to the
CDM$\Lambda$ runs ($\Omega_0=0.25$, $\Lambda=0.75$). The solid line
shows the ``natural'' scaling, $\delta_c \propto \Omega_0 (1+z)^3$
expected if the characteristic density of a halo is directly
proportional to the mean matter density of the universe at the time of
collapse.}
\end{figure}

\begin{figure}
%\plotone{mrhoc_c.ps}
\plotone{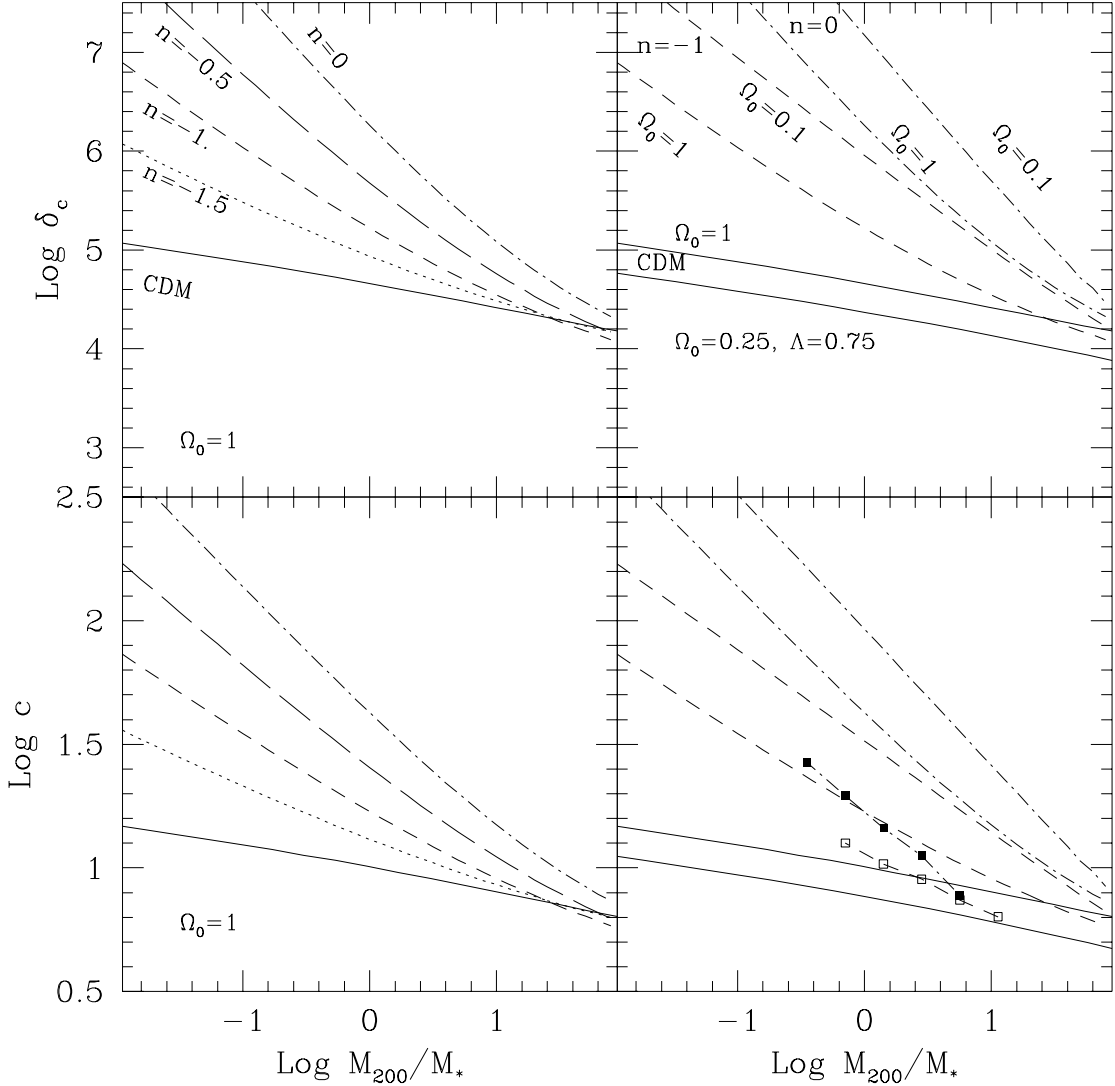}
\figurenum{9}
\caption{
A summary of the mass dependence of $\delta_c$ and $c$ in our
different cosmological models.  Mass is given in units of the
nonlinear mass scale $M_{\star}$.  Curves are labeled in the upper
plots and the same line types are used in the bottom plots. The
symbols in the lower-right panel show the correlation between $c$ and
mass found by Cole \& Lacey (1996), with open squares for $n=-1$,
$\Omega_0=1$ and solid squares for $n=0$, $\Omega_0=1$. These results
should be compared to the lower dashed and lower dot-dashed curves in
the same panel, respectively. Because of their poorer numerical
resolution, Cole \& Lacey's halos are significantly less concentrated
than the ones in our study.}
\end{figure}

\begin{figure}
%\plotone{scatterzf_new.ps}
\plotone{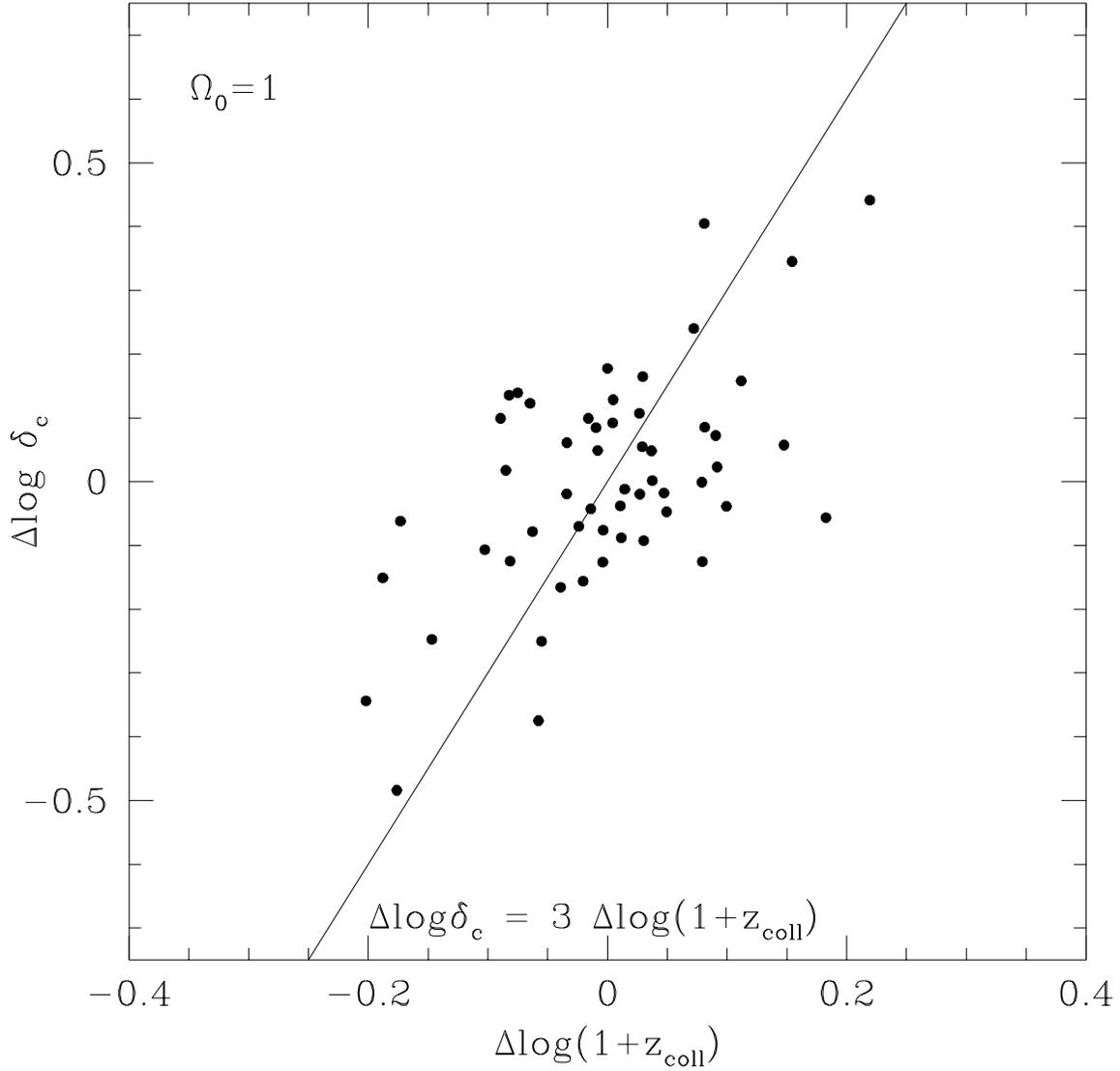}
\figurenum{10}
\caption{
The $\delta_c$-deviation from the fits shown in Figure 5 (for $f=0.01$)
plotted versus the deviation from the mean relation between the
collapse redshift, $1+z_{coll}$ (Figure 8), and the mass of a
system. Note that systems assembled earlier (later) than the average
tend to have characteristic densities above (below) the mean. The
correlation between the $\delta_c$ and $1+z_{coll}$ residuals follows
the ``natural'' scaling, $\Delta \log \delta_c = 3 \Delta \log
(1+z_{coll})$, as shown by the solid line. Only the results
corresponding to $\Omega_0=1$ are shown. }
\end{figure}

\begin{figure}
%\plotone{scatterlmb_new.ps}
\plotone{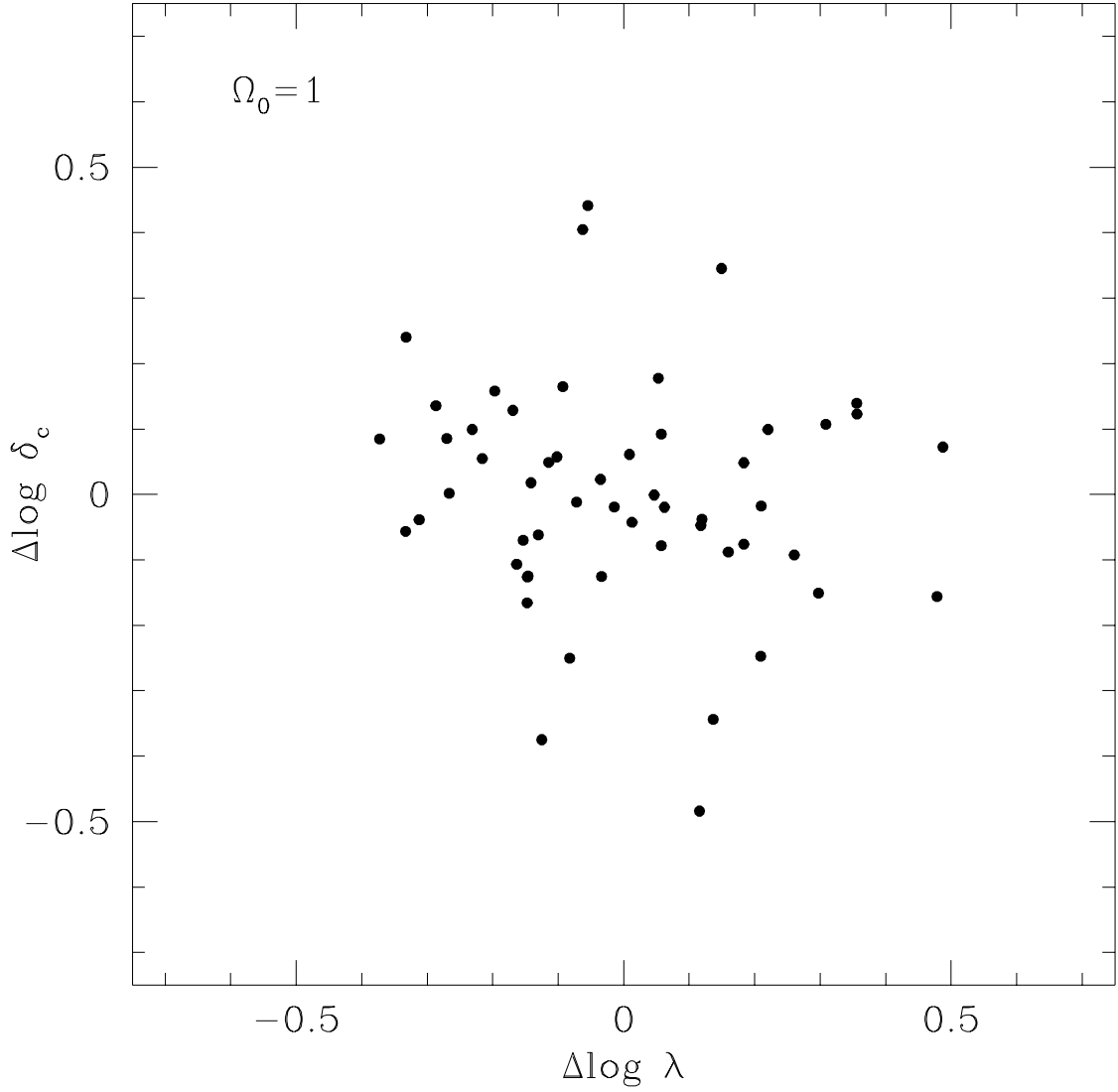}
\figurenum{11}
\caption{
The $\delta_c$-residuals in Figure 5 relative to the solid curve fit
($f=0.01$) plotted versus the residuals in the $M_{200}$--$\lambda$
correlation. This figure shows that the scatter in $\delta_c$ at a
given mass cannot be attributed to the effects of rotation. All halos
with $\Omega_0=1$ are included in this plot. }
\end{figure}

\begin{figure}
%\plotone{zurek_1.ps}
\plotone{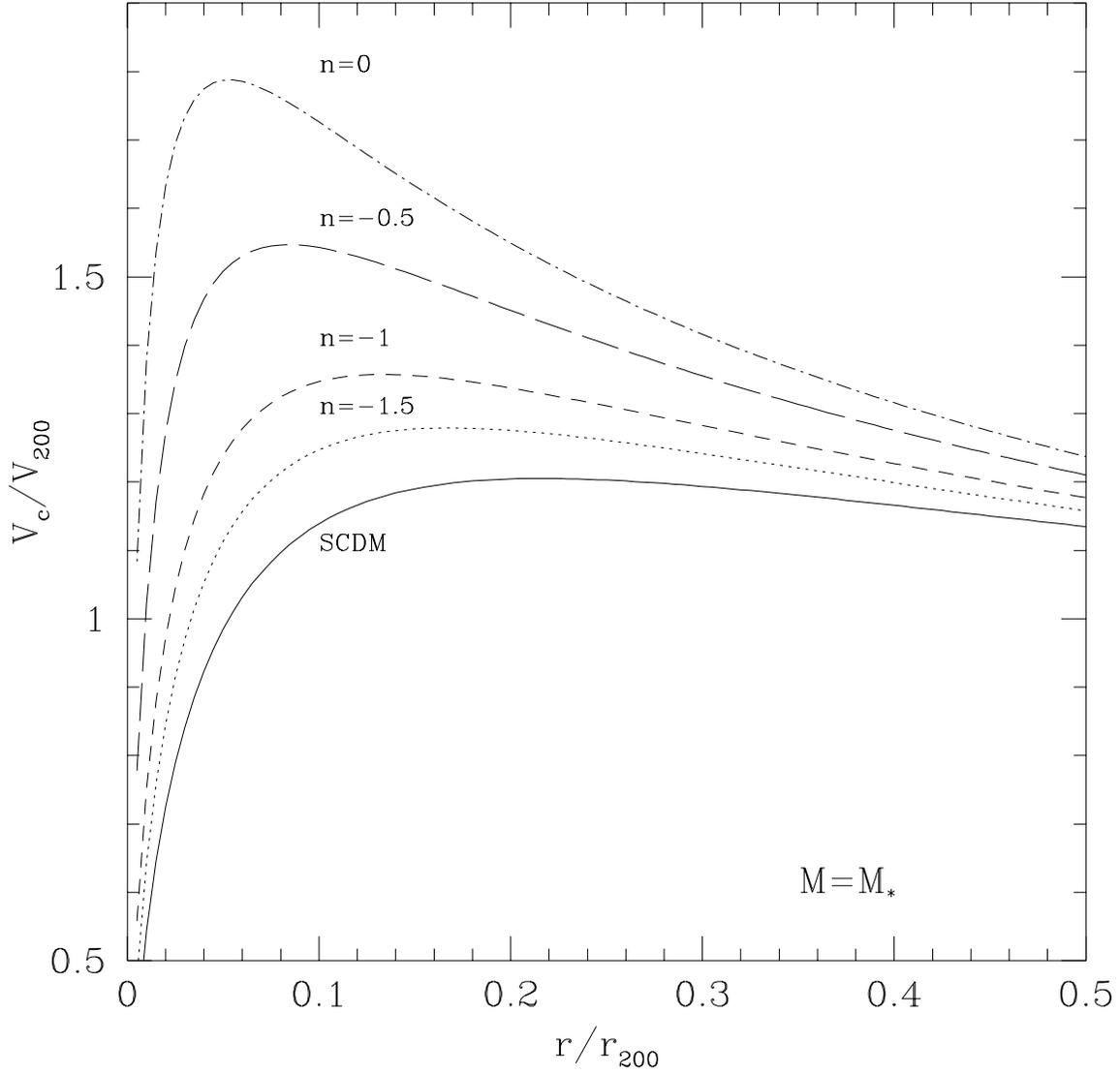}
\figurenum{12}
\caption{
Circular velocity profiles of $M_{\star}$ halos formed in
simulations with different power spectra in an Einstein-de Sitter
universe. The curves were computed using eq.~3 and the values of the
concentration $c$ obtained from the fits in Figure~9. Radii are in units of the virial radius and circular speeds
in units of the circular velocity at the virial radius. Note that,
because of the use of linear units, the fact that all curves have the
same shape (eq.~3) is not immediately apparent.}
\end{figure}

\begin{figure}
%\plotone{crone.ps}
\plotone{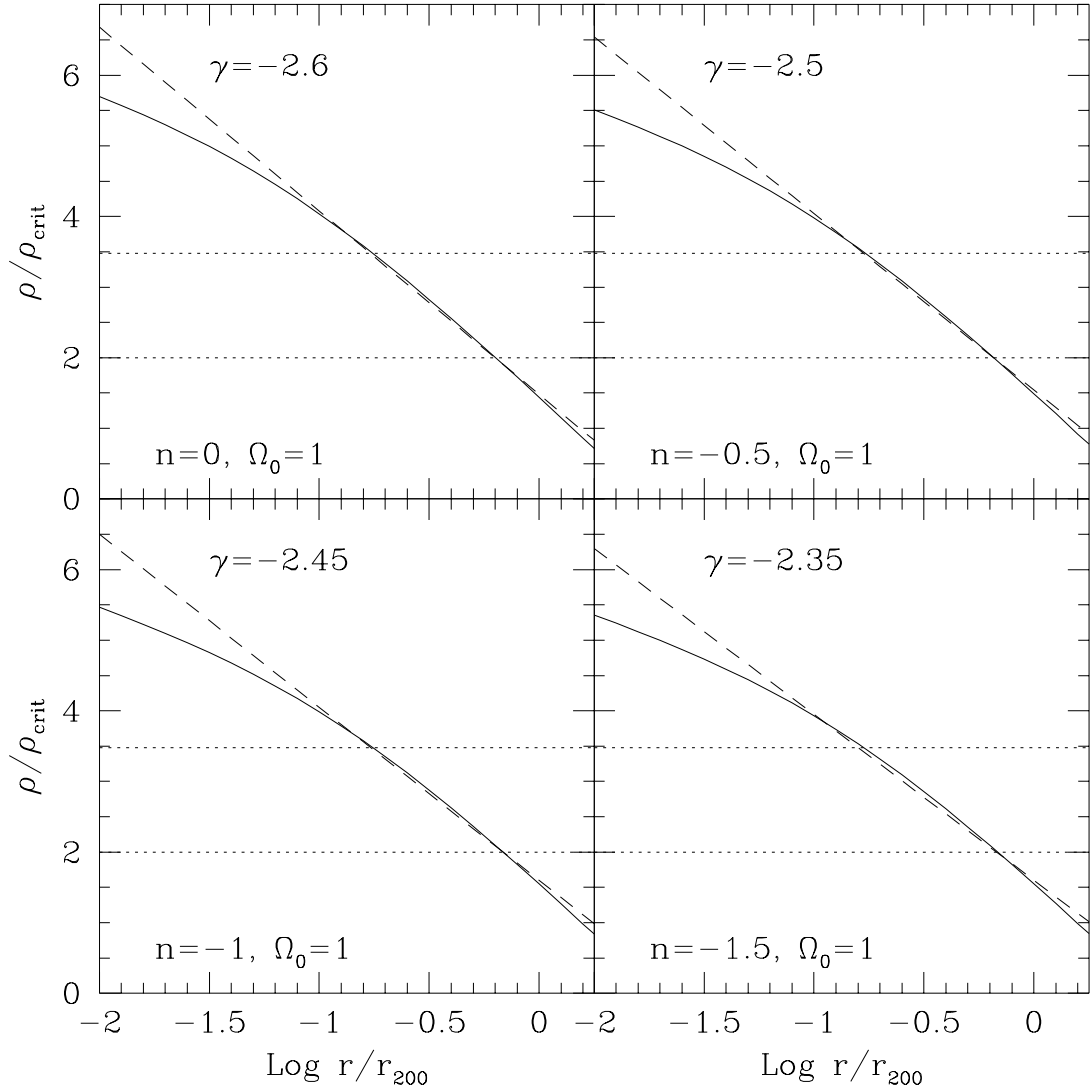}
\figurenum{13}
\caption{ A comparison between our density profiles and the power-law fits
of Crone et al. (1994). The region fitted by these authors, corresponding
to densities between $100$ and $3000$, is shown by the dotted lines. Over
this region our results (solid lines) and the power-law parameterization
adopted by Crone et al. (dashed lines) are consistent. The profiles shown
by the solid curves correspond to $M=6 M_{*}$ ($n=-1.5$), $M=4M_{*}$
($n=-1$), $M=2M_{*}$ ($n=-0.5$), and $M=M_{*}$ ($n=0$), as discussed in the
text. } 
\end{figure} 

%\begin{figure} 
%%\plotone{dwarfs.ps} 
%\figurenum{1}
%\caption{ 
%The circular velocity profiles of CDM halos compared with the rotation 
%curves of four dwarf irregulars. References for the observational data 
%are given in the text. All CDM halos have been chosen to match the
%maximum of the rotation curve. The solid and dotted curves correspond 
%to the SCDM and CDM$\Lambda$ models, respectively. The dashed curves
%correspond to a lower density CDM model ($\Omega_0=0.1$, $\Lambda=0.9$)
%normalized to COBE. Low values of $\Omega$ are favored by these
%observations.}  
%\end{figure} 

% That's all, folks.  
% 
\end{document}